\documentclass[11pt]{article} 

\usepackage{amsmath,amsthm,latexsym,amssymb,amsfonts,epsfig}

\usepackage{float}
\usepackage{cite}
\usepackage{hyperref}
\hypersetup{
    colorlinks,%
    citecolor=blue,%
    filecolor=blue,%
    linkcolor=blue,%
    urlcolor=blue
}

\newcommand{\be}{\begin{equation}}
\newcommand{\ee}{\end{equation}}
\newcommand{\ba}{\begin{eqnarray}}
\newcommand{\ea}{\end{eqnarray}}

\title{{\sf Hamiltonian Renormalization III.}\\ 
{\sf Renormalisation Flow of 1+1 dimensional free scalar fields:}\\ 
{\sf Properties}}
\author{
{\sf T. Lang}$^1$\thanks{{\sf 
thorsten.lang@gravity.fau.de}},
{\sf K. Liegener}$^1$\thanks{{\sf 
klaus.liegener@gravity.fau.de}},
{\sf T. Thiemann}$^1$\thanks{{\sf 
thomas.thiemann@gravity.fau.de}}\\
\\
{\sf $^1$ Inst. for Quantum Gravity, FAU Erlangen -- N\"urnberg,}\\
{\sf Staudtstr. 7, 91058 Erlangen, Germany}\\
}
\date{{\small\sf \today}}

\makeatletter
\@addtoreset{equation}{section}
\makeatother

\begin{document}

\maketitle

{\sf
\begin{abstract}
\fontfamily{lmss}\selectfont{
This is the third paper in a series of four in which a renormalisation flow is introduced which
acts directly on the Osterwalder-Schrader data (OS data) without recourse to a path integral.
Here the OS data consist of a Hilbert space, a cyclic vacuum vector therein and a Hamiltonian  annihilating the vacuum which can be obtained from an OS measure, that is a measure respecting
(a subset of) the OS axioms.

In the previous paper we successfully tested our proposal for the two-dimensional massive Klein-Gordon model, that is, we could confirm that our framework finds the correct fixed point
starting from a natural initial  naive discretisation of the finite resolution Hamiltonians, in particular
the underlying Laplacian on a lattice, and a natural coarse graining map that 
drives the renormalisation flow. However, several questions remained unanswered.
How generic can the initial  discretisation and the coarse graining map be in order that the 
fixed point is not changed or is at least not lost, in other words, how universal is the fixed
point structure? Is the fixed point in fact stable, that is, is the fixed point actually a limit of the 
renormalisation sequence?  We will address these questions in the present paper. 
}\end{abstract}

\newpage

\tableofcontents

\newpage
~\\
{\bf Notation:}\\
\\
In this paper we will deal with quantum fields in the presence of an
infrared cut-off $R$ and with smearing functions of finite time support
in $[-T,T]$. The spatial ultraviolet cut-off is denoted by $M$ and has the
the interpretation of the number of lattice vertices in each spatial
direction. We will mostly not be interested in an analogous temporal
ultraviolet cut-off $N$ but sometimes refer to it for comparison
with other approaches. These quantities allow us to define dimensionful
cut-offs $\epsilon_{RM}=\frac{R}{M},\; \delta_{TN}=\frac{T}{N}$. In
Fourier space we define analogously
$k_R=\frac{2\pi}{R},\; k_M=\frac{2\pi}{M},\;
k_T=\frac{2\pi}{T},\; k_N=\frac{2\pi}{N}$.

We will deal with both instantaneous fields, smearing functions and Weyl
elements as well as corresponding temporally dependent objects.
The instantaneous objects are denoted by lower case letters
$\phi_{RM}, \; f_{RM},\; w_{RM}[f_{RM}]$, the temporally dependent ones by
upper case ones
$\Phi_{RM}, \; F_{RM},\; W_{RM}[F_{RM}]$. As we will see, smearing functions
$F_{RM}$ with
compact and discrete (sharp) time support will play a more fundamental role
for our purposes than those with a smoother dependence.

Osterwalder-Schrader reconstruction concerns the interplay between
time translation invariant, time reflection invariant and reflection positive
measures (OS measures)
$\mu_{RM}$ on the space of history fields $\Phi_{RM}$ and their corresponding
Osterwalder-Schrader (OS) data ${\cal H}_{RM}, \Omega_{RM}, H_{RM}$ where
${\cal H}_{RM}$ is a Hilbert space with cyclic (vacuum) vector $\Omega_{RM}$
annihilated by a self-adjoint Hamiltonian $H_{RM}$. Together, the vector
$\Omega_{RM}$ and the scalar product $<.,.>_{{\cal H}_{RM}}$ define
a measure $\nu_{RM}$ on the space of instantaneous fields $\phi_{RM}$.

Renormalisation consists in defining a flow or sequence
$n\to \mu^{(n)}_{RM},\; n\in \mathbb{N}_0$ for all
$M$ of families of measures $\{\mu^{(n)}_{RM}\}_{M\in \mathbb{N}}$.
The flow will be defined in terms of a coarse graining or embedding
map $I_{RM\to M'},\; M<M'$ acting on the smearing functions and satisfying
certain properties that will grant that 1. the resulting fixed point family
of measures, if it exists, is cylindrically consistent and 2. the flow
stays within the class of OS measures. Fixed point quantities are denoted
by an upper case $^\ast$, e.g. $\mu^\ast_{RM}$.

\newpage

\section{Introduction}
\label{Introduction}
\numberwithin{equation}{section}

In this article we continue our study of a test model, started in our companion paper \cite{LLT2},  
for a direct Hamiltonian renormalisation flow of Osterwalder-Schrader data (OS data) derived in our companion paper \cite{LLT1} 
which does not need any intermediate recourse to a path integral construction. Developed in the framework of constructive QFT \cite{GJ87,WV90,Bal84,GK80,JLMSSGH84,BFS79,Sei82,BY76}, the OS data obtained from a path integral measure satisfying the OS axioms \cite{OS72} (or a sufficient subset thereof\cite{AMMT99})
via OS reconstruction  consist of a Hilbert space, a cyclic vector therein for the algebra of smeared sharp time zero
(configuration) fields and a Hamiltonian annihilating that vacuum. Conversely, as we showed in 
\cite{LLT1}, the OS measure that gives rise to given OS data can be chosen as the Wiener 
(or heat kernel) measure that is defined by them \cite{GJ87}. To have access to a renormalisation
flow directly of the OS data is of tremendous practical importance because it is much easier 
to compute the matrix elements of a given Hamiltonian than its corresponding heat kernel 
(i.e. the matrix elements of the corresponding contraction semigroup) especially in the interacting
case. 

In \cite{LLT2} we showed that the fixed point of the direct flow and that of the more traditional
Wilsonian path integral flow define the same continuum QFT, 
at least for the 1+1-dimensional free scalar QFT. Moreover, the direct flow gives one much 
more direct access to the approximate finite resolution matrix elements of the continuum 
Hamiltonian, where the approximation quality is labelled by the number of renormalisation
(block spin transformation) steps. This result can be considered as a signal that the 
renormalisation flow defined is a physically viable one.

What is missing in that analysis is the investigation of stability and universality  questions: Is the fixed point found actually a limit and thus a stable fixed point or just an accumulation point? How much does the value or the existence of the fixed point depend on the chosen initial OS data and 
on the chosen coarse graining map that defines the renormalisation flow? How does 
the renormalised Laplacian look like at finite resolution, i.e. how do contributions from 
far separated lattice points decay?

We address these questions in the present paper whose architecture is as follows:\\
\\
In section \ref{s2} we briefly review some of the ingredients of \cite{LLT1, LLT2} for the 
convenience of the reader. For more details, the reader is referred to those papers.

In section \ref{s3} we analyse the convergence behaviour, stability properties and 
universality features of the Hamiltonian flow defined.

In section \ref{s4} we study the perfect lattice Laplacian with respect to the n-th order 
neighbour contributions.

In section \ref{s5} we give an outlook into the regime of interacting fields, specifically
gauge fields whose initial finite resolution Hamiltonian is more naturally discretised 
using Wilson like (holonomy) variables \cite{WK73,KS75}. We also summarise our findings.

In the appendix we review and compare to standard coarse graining maps.\cite{WK73, Wil75} \\
\\
This article is followed by the fourth paper \cite{LLT4} in this series in which we remove the restriction to two spacetime dimensions which enables us to address the issue of detecting
at finite resolution, that is, although one considers a finite lattice, whether the flow approaches
a rotationally invariant theory. This serves as a basic example for the 
preservation or rather restoration of more general (in particular gauge), symmetries that are   
 broken by naive discretisations \cite{BD09(Broken)} and which are a strong motivation for this entire series.

\section{Review of direct Hamiltonian Renormalisation}
\label{s2}
This section serves to recall the notation and elements of Hamiltonian 
renormalisation from \cite{LLT1, LLT2} to which the reader is referred to for all the details.\\
\\
We consider infinite dimensional conservative Hamiltonian systems on globally hyperbolic 
spacetimes of the form $\mathbb{R}\times \sigma$.  If  $\sigma$ is not compact
 one introduces an infrared (IR) cut-off $R$ for the spatial manifold $\sigma$ by restricting to test-functions which are defined on a compact submanifold, e.g. a torus $\sigma_R:=[0,R]^D$ if $\sigma=\mathbb{R}^D$. We will assume this cut-off $R$ to be implicit in all formulae 
 below but do not display it to keep them simple, see \cite{LLT1,LLT2} for the explicit
 appearance of $R$.

Moreover, an ultraviolet (UV) cut-off $\epsilon_M:=R/M$ is introduced by restricting the smearing functions $f_M$ to a finite spatial resolution. In other words, $f_M\in L_{M}$ is defined by its value on the vertices of a graph, which we choose here to be a cubic lattice, i.e. there are $M^D$ many vertices, labelled by $m\in\mathbb{Z}^D_M$, $\mathbb{Z}_M=\{0,1,...,M-1\}$. In this 
paper we consider a background dependent (scalar) QFT and thus we have access to a natural 
inner product defined by it. For more general theories this structure is not available
but the formalism does not rely on it. The scalar products for $f_M,g_M\in L_{M}$ and respectively for $f,g\in L=C^{\infty}([0,R])$ are defined by
\begin{align}
\langle f_M, g_M \rangle_M = \epsilon^D_M\sum_{m\in\mathbb{Z}^D_M} \bar{f}_M(m) g_M(m), \hspace{20pt}
\langle f,g \rangle =\int_{\sigma_R} d^Dx \bar{f}(x)g(x)
\end{align} 
where $\bar{f}$ denotes the complex conjugate of $f$. \\
Given an $f_M:\mathbb{Z}^D_M\rightarrow \mathbb{R}$ we can embed it into the continuum by  an {\it injection map} $I_M$
\begin{equation} \label{injection map}
\begin{split}
I_M : L_{M} &\rightarrow L \\
  f_M &\mapsto (I_M f_M ) (x):=\sum_{m\in \mathbb{Z}^D_M}f_M(m)\chi_{m\epsilon_M}(x)
\end{split}
\end{equation}
Note that indeed the coefficient $f_M(m)$ is the value of $I_M f_M$ at $x=m\epsilon_M$.

$L$ is much larger than the range of $I_M$. This allows us to define its corresponding left inverse: {\it the evaluation map} $E_M$ is found to be
\begin{equation} \label{evaluation map}
\begin{split}
E_M : L &\rightarrow L_{M} \\
  f &\mapsto (E_M f ) (m):=f(m\epsilon_M)
\end{split}
\end{equation}
and by definition obeys
\begin{align}\label{left-inverse}
E_M\circ I_M= \text{id}_{L_M}
\end{align}
Given those maps we are now able to relate test functions and thus also observables from the continuum with their discrete counterpart, e.g. for a smeared scalar field one defines: 
\begin{align}
\phi_M[f_M]:=\langle f_M, \phi_M\rangle_M,\hspace{20pt} \phi_M(m):=(I_M^\dagger \phi)(m)=\int_{[0,R)^D}d^Dx\;\chi_{m\epsilon_M}(x)\phi(x)
\end{align}
Indeed, since the kernel of any map $C: L\rightarrow L$ in the continuum is given as
\begin{align}
\langle f, C g\rangle =: \int_{[0,R]^D}d^Dx\int_{[0,R]^D}d^Dy\; C(x,y)\bar{f}(x)g(y)
\end{align}
it follows
\begin{align}
\langle I_M f_M, C I_M g_M\rangle = \langle f_M, [I^{\dagger}_M C I_M]g_M\rangle_M=: \langle f_M, C_M g_M\rangle_M
\end{align}
which shows that
\begin{align}
C_M(m,m')=\epsilon^{-2D}_M\langle \chi_{m\epsilon_M},C \chi_{m'\epsilon_M}\rangle
\end{align}
\\
with $\chi_{m\epsilon_M}(x):=\prod_{a=1}^D\chi_{[m^a\epsilon_M,(m^a+1)\epsilon_M)}(x)$.\\
The concatenation of evaluation and injection for different discretisations shall be called {\it coarse graining map} $I_{M\rightarrow M'}$ if $M<M'$:
\begin{align}
I_{M\rightarrow M'} = E_{M'} \circ I_M : L_{R,M} \rightarrow L_{R,M'}
\end{align}
as they correspond to viewing a function defined on the coarse lattice as a function on a finer lattice of which the former is not necessarily a sublattice although. In practice we will 
choose the set of $M$ such that it defines a partially ordered and directed index set.
The coarse graining map is a free choice of the renormalisation process whose flow it drives and its viability can be tested only a posteriori if we found a fixed pointed theory which agrees with the measurements of the continuum. Hence proposals for such a map should be checked at least in simple toy-models before one can trust their predictions.

The coarse graining maps are now used to call a family of 
of measures $M\mapsto \nu_M$ on a suitable space of fields $\phi_M$ {\it cylindrically 
consistent} iff 
\begin{align}\label{cylindricalconsistency}
\nu_{M}(w_M[f_M])=\nu_{M'}(w_{M'}[I_{M\rightarrow M'}\circ f_M])
\end{align}
where $w_M$ is a Weyl element restricted to the configuration degrees of freedom, i.e. for a 
scalar field theory as in the present paper 
$w_M[f_M]=\exp(i\phi_M[f_M])$. The measure $\nu_M$  can be considered as the positive linear 
GNS functional on the Weyl $^\ast-$algebra generated by the $w_M[f_M]$ with GNS data 
$({\cal H}_M,\Omega_M)$, that is
\begin{align}
\nu_M(w_M[f_M])=\langle \Omega_M , w_M[f_M]\Omega_M\rangle_{{\cal H}_M}
\end{align}
In particular, the span of the $w_M[f_M]\Omega_M$ lies dense in $\mathcal{H}_M$ and 
we simplify the notation by refraining from displaying a possible GNS null space. 
Under suitable conditions \cite{Yam75} a cylindrically consistent family has a continuum measure $\nu$ as a projective limit
which is related to its family members $\nu_M$ by 
\begin{align} \label{projectivelimit}
\nu_M(w_M[f_M])=\nu(w[I_M f_M]) 
\end{align}
It is easy to see that (\ref{projectivelimit}) and (\ref{cylindricalconsistency}) are compatible
iff we constrain the maps $I_M, E_M$ by the condition for all $M<M'$
\begin{align}\label{cylconrel}
I_{M'}\circ I_{M\rightarrow M'} = I_M
\end{align}
This constraint which we also call {\it cylindrical consistency} means that injection into the continuum 
can be done independently of the lattice on which we consider the function to 
be defined, which is a physically plausible assumption. 

 In the language of the GNS data  $(\mathcal{H}_M,\Omega_M)$ cylindrical consistency means that 
 the maps 
 \begin{align}
J_{M\rightarrow M'} : \mathcal{H}_M & \rightarrow \mathcal{H}_{M'}\\
w_M[f_M]\Omega_M & \mapsto w_{M'}[I_{M\rightarrow M'} f_M]\Omega_{M'}
\end{align}
define a family of {\it isometric} injections of Hilbert spaces. The continuum GNS data 
are then given by the corresponding inductive limit, i.e. the embedding of Hilbert spaces 
defined densely by $J_M w_M[f_M]\Omega_M=w[I_M f_M] \Omega$ which is isometric. 
Note that $J_{M'} J_{M\rightarrow M'}=J_M$. 
The GNS data are completed by a family of positive 
self-adjoint Hamiltonians $M\mapsto H_M$ defined 
densely on the $w_M[f_M]\Omega_M$ to the OS data $({\cal H}_M, \; \Omega_M,\; H_M)$.
We define a family of such Hamiltonians to be cylindrically consistent provided that
\begin{align}
J_{M\rightarrow M'}^\dagger H_{M'} J_{M\rightarrow M'}=H_M
\end{align}
It is important to note that this does {\it not} define an inductive system of operators which 
would be too strong to ask and thus 
does not grant the existence of a continuum Hamiltonian. However, it grants the existence 
of a continuum positive quadratic form densely defined by 
\begin{align}
J_M^\dagger H J_M =H_M
\end{align}
If the quadratic form can be shown to be closable, one can extend it to a positive self-adjoint 
operator.    

In practice, one starts with an {\it initial} family of OS data $({\cal H}^{(0)}_M,\;
\Omega^{(0)}_M,\; H^{(0)}_M)$ usually obtained by some {\it naive} discretisation 
of the classical Hamiltonian system and its corresponding quantisation. The corresponding
GNS data will generically not define a cylindrically consistent family of measures, i.e. 
the maps $J_{M\rightarrow M'}$ will fail to be isometric. Likewise, the family 
of Hamiltonians will generically fail to be cylindrically consistent. Hamiltonian 
renormalisation now consists in defining a sequence of an {\it improved} OS data 
family $n\mapsto 
 ({\cal H}^{(n)}_M,\; \Omega^{(n)}_M,\;H^{(n)}_M)$ defined inductively by 
 \begin{align} \label{improving}
J^{(n)}_{M\rightarrow M'} w_M[f_M] \Omega^{(n+1)}_M:=w_{M'}[I_{M\rightarrow M'} f_M]
\Omega^{(n)}_{M'},\;\;H^{(n+1)}_M:=J^{(n)}_{M\rightarrow M'} H^{(n)}_{M'} J^{(n)}_{M\rightarrow M'}
\end{align}
Note that $H^{(n)}_M \Omega^{(n)}_M=0$ for all $M,n$. 
If the corresponding flow (sequence) has a fixed point family $({\cal H}^\ast_M,\; \Omega^\ast_M,\;
H^\ast_M)$ then its internal cylindrical consistency is restored by construction.

\section{Properties of the renormalisation sequence}
\label{s3}

Recall from \cite{LLT2} that the flow of the covariance $c^{(n)}_M$ of the measure 
$\nu^{(n)}_M$ for the 1+1-dimensional Klein-Gordon field of mass $p$ is given 
by the following set of formulae (up to numerical prefactors which are not important 
for what follows)
\begin{eqnarray}
\nu^{(n)}_M(w_M[f_M]) &=& e^{-\frac{1}{2} \langle f_M, c^{(n)}_M f_M\rangle_{M}}
\nonumber\\
\langle f_M, c^{(n)}_M f_M\rangle_{M} &:=& \epsilon_M^2 \sum_{m,m'\in \mathbb{Z}_M} 
\bar{f}(m) \; c^{(n)}_M(m,m')\; f_M(m')
\nonumber\\
c^{(n)}_M(m,m'):&=& c^{(n)}_M(m-m')= \int_{\mathbb{R}} \; \frac{dk_0}{2\pi}\;
\sum_{l\in \mathbb{Z}_M}\; e^{ik_M l(m-m')}\;
\hat{c}^{(n)}_M(k_0,l)
\nonumber\\
\hat{c}^{(n+1)}_M(k_0,l)&=& \frac{1}{2} [(1+\cos(l k_M/2)\hat{c}^{(n)}_{2M}(k_0,l)+
(1-\cos(l k_M/2)\hat{c}^{(n)}_{2M}(k_0,l+M)]
\end{eqnarray}
where $k_M:=\frac{2\pi}{M}$. If we start the recursion with  the naive discretisation $(\Delta^{(0)}_M f_M):=\epsilon_M^{-2}
[f_M(m+1)+f_M(m-1)-2f_M(m)]$ of the Laplacian then it turns out that the recursion can 
be parametrised by three functions $a_n,\; b_n,\; c_n$ of $q_M:=\sqrt{k_0^2+p^2}\epsilon_M$
\begin{align}
\hat{c}^{(n)}(k_0,l)=
\frac{\epsilon_M^2}{q_M^3}\; \frac{b_n(q_M)+c_n(q_M) \cos(t_M)}{a_n(q_M)-cos(t_M)}
\end{align}
with $t_M=l k_M$. The flow is then defined 
in terms of the recursion relations for the parametrising functions
\begin{align}
a_{n+1}(q)&:=2 [a_n(q/2)]^2-1,\nonumber\\
b_{n+1}(q)&:=[2a_n b_n+b_n+c_n+ a_n c_n](q/2),\nonumber\\
c_{n+1}(q)&:=2[b_n+c_b+ a_n c_n](q/2)
\end{align}
with the initial values 
\begin{align}
a_0(q)=1+q^2/2,\;b_0(q)=q^3/2,\; c_0(q)=0
\end{align}
corresponding to the above chosen naive discretisation of the Hamiltonian. 
With these initial values, the recursion has the unique fixed point 
\begin{align}
a^\ast(q)={\rm ch}(q),\; b^\ast(q)=q{\rm ch}(q)-{\rm sh}(q),\;c^\ast(q)={\rm sh}(q)-q
\end{align}

What we did not check in \cite{LLT2} is whether this fixed point is actually a limit of 
the recursion or merely an accumulation point. Moreover, the stability of the fixed point 
with respect to perturbing the initial values was not considered. In what follows we 
supply parts of this analysis.

\subsection{Convergence properties}
\label{convergence}

A necessary condition for convergence of the flow is that 
\begin{align}
a_n(q)=2a_{n-1}(q/2)^2 -1
\end{align}
with starting value $a_0(q) = 1+\frac{1}{2}q^2$ really runs into its fixpoint
\begin{align}
\cosh (q) =\sum_n \frac{1}{(2n)!}q^{2n}=1+\frac{1}{2}q^2+\frac{1}{24}q^4+...
\end{align}
We will examine this by computing the flow of the coefficient 
of each power of $q^2$ separately. This maps the problem of dealing with recursive 
functional equations to recursive relations of sequences, see e.g. \cite{KN12,San90}. One immediately 
sees that the constant and the quadratic term always remain the same under the flow, i.e. $a^{(0)}_n=1$, $a^{(2)}_n=\frac{1}{2}$ for all $n$ where $a_n(q)=\sum_{k=0}^{2^{n+1}} 
a^{(k)}_n q^k$. In fact, it is  easy to see that all odd powers of $q$ vanish and that $a_n$ 
is a polynomial of order $2^n$ in $q^2$.     
For the remaining coefficients we note the following lemmata:\\
{\bf Lemma 1:} Suppose $f,g\in \mathbb{R},\;f\not= 1$. Then for a sequential recursive relation of the form
\begin{align}
a_n = f a_{n-1}+g
\end{align}
we find a solution as:
\begin{align}
a_n=f^n (a_0 -\frac{g}{1-f})+\frac{g}{1-f}
\end{align}
\begin{proof}
It is obviously true for $n=1$ as $a_1= fa_0+\frac{g}{1-f}(1-f)=fa_0+g$. Assuming thus the claim holds for $n$ it follows
\begin{align}
a_{n+1}=f\left(f^n(a_0-\frac{g}{1-f})+\frac{g}{1-f}\right)+g =f^{n+1}(a_0-\frac{g}{1-f})+\frac{g}{1-f}
\end{align}
\end{proof}
The lemma can be extended to $f=1$ using de l'Hospital's theorem.\\
{\bf Lemma 2} Suppose that
$f(n), g(n)$ are sequences with $f(n)\neq 0 \forall n$. Then a sequential 
recursive relation of the form:
\begin{align}
a_{n+1} = f(n) a_n +g(n)
\end{align}
is solved by
\begin{align}
a_n=\left(\prod_{k=0}^{n-1}f(k)\right)\left(a_0+\sum_{j=0}^{n-1}\frac{g(j)}{\prod_{k=0}^jf(k)}\right)
\end{align}
\begin{proof}
Let $A_n:=a_n/(\prod_{k=0}^{n-1}f(k)),\; n\ge 1$ and $A_0:=a_0$. Then by the recursion relation
\begin{align}
A_{n+1}-A_n=\frac{g(n)}{\prod^n_{k=0}f(k)}
\end{align}
hence 
\begin{align}
A_n-A_0=\sum_{j=0}^{n-1}A_{j+1}-A_j=\sum_{j=0}^{n-1}\frac{g(j)}{\prod_{k=0}^jf(k)} \Rightarrow a(n)=\left(\prod^{n-1}_{k=0}f(k)\right)\left(A_0+\sum_{j=0}^{n-1}\frac{g(j)}{\prod^j_{k=0}f(k)}\right)
\end{align}
\end{proof}
It is instructive to verify that Lemma 2 reduces to Lemma 1 when $f(n),g(n)$ do 
not depend on $n$. 

Since $a_0(q)$ is quadratic and the recursion is quadratic as well, we see that the highest power for $a_n(q)$ is always $2^{n+1}$. Now, we apply the Cauchy-product-rule
\begin{align}
\left(\sum_{k=0}^{\infty}a_kq^k\right)\left(\sum_{k=0}^{\infty}b_kq^k\right)=\sum_{k=0}^{\infty}\left(\sum_{j=0}^ka_{k-j}b_j\right) q^k
\end{align}
to get (using that all the coefficients of all odd powers vanish)
\begin{align}
a_{n+1}(q)&=\sum_{k=0}^{2^{n+2}}a^{(k)}_{n+1}q^{k}= 2 \left(\sum_{k=0}^{2^{n+1}}a^{(k)}_n \left(\frac{q}{2}\right)\right)^2-1=2\sum_{k=0}^{2^{n+2}}\left(\sum_{j=0}^ka_n^{(j)}a^{(k-j)}_n\right)2^{-k}q^k-1
\end{align}
So as long as $k\geq 4$:
\begin{align}
a^{(k)}_{n+1}&=\sum^k_{j=0}a^{(j)}_na^{(k-j)}_n2^{-k+1}=2^{-k+2}a^{(k)}_n+2^{-k+1}\sum^{k-1}_{j=1}a^{(j)}_na_n^{(k-j)}
\end{align}
which is now a linear relation for $a^{(k)}_n$, i.e. we can apply Lemma 2. Now since all $a^{(k)}_l =0$, $\forall l \leq \lfloor \frac{\ln (k/2)}{\ln 2}\rfloor$ also our starting value is zero and we get:
\begin{align}
a_m^{(k)}&=(2^{-k+2})^m\sum_{t=0}^{m-1}(2^{-k+2})^{-(t+1)}\left(2^{-k+1}\sum^{k-1}_{j=1}a^{(j)}_t a^{(k-j)}_t\right)=\sum^{m-1}_{t=0}\frac{1}{2}(2^{-k+2})^{m-t}\sum^{k-1}_{j=1}\left(a^{(j)}_ta_t^{(k-j)}
\right)
\end{align}
It is easy to compute, e.g.:
\begin{align}
a^{(4)}_n=\frac{1}{4!}(1-2^{-2n}) 
\end{align}
and use this to claim
\begin{align}\label{falloff-for-a}
a^{(2k)}_n=\frac{1}{(2k)!}\left(1+\mathcal{O}(2^{-n})\right), a^{(2k+1)}_n=0
\end{align}
which is the above equation for $k\leq 2$. And assuming it holds for $\forall j \leq k$
\begin{align} \label{order}
a^{(2k)}_m&=(2^{-2k+2})^m\sum_{t=0}^{m-1}(2^{-2k+2})^{-t}\frac{1}{2}\sum_{j=1}^{2k-1}a^{(j)}_ta^{(2k-j)}_t=2^{2(1-k)m}\sum_{t=0}^{m-1}2^{2(k-1)t}\frac{1}{2}\sum_{j=1}^{k-1}a^{(2j)}_ta^{(2k-2j)}_t=\nonumber\\
&=\sum^{m-1}_{t=0}2^{2(1-k)m-t}\frac{1}{2}\sum_{j=1}^{k-1}\frac{1}{(2j)!(2k-2j)!}\left(1+\mathcal{O}(2^{-t})\right)\left(1+\mathcal{O}(2^{-t})\right)
\end{align}
We now expand the function of $2^{-t}$ appearing in (\ref{order}) 
as a power series $\sum_i c_i (2^{-t})^i$ with some coefficients $c_i\in\mathbb{R}$, such that the $c_i$ are independent of $t$ and finite. It follows:
\begin{align}
a^{(2k)}_m
&=\frac{1}{(2k)!}\sum_{j=1}^{k-1}\frac{(2k)!}{(2j)!(2k-2j)!}\frac{1}{2}2^{2(1-k)m}\left(
\sum_{t=0}^{m-1}2^{2(k-1)t}+\sum_ic_i (2^{-t})^i\right)=\nonumber\\
&=\frac{1}{(2k)!}\left(\sum_{j=0}^k\left(\begin{array}{c}
2k\\
2j
\end{array}\right)-2\right)
2^{2(1-k)m-1}\left(\frac{1-2^{2(k-1)m}}{1-2^{2(k-1})}+\sum_ic_i\frac{1-2^{-im}}{1-2^{-i}}\right)\nonumber\\
&=\frac{1}{(2k)!}\left(\sum_{j=0}^{2k-1}\left(\begin{array}{c}
2k-1\\
j
\end{array}\right)-2\right)
\frac{2^{-m2(k-1)-1}}{2^{2(k-1)}-1}\left(
2^{2(k-1)m}-1+\sum_ic_i\frac{2^{2(k-1)}-1}{1-2^{-i}}(1-2^{-mi})
\right)
\nonumber\\
&=\frac{1}{(2k)!}(2^{2k-1}-1)\frac{1}{2^{2(k-1)}-1}\left(1-(2^{-m})^{2(k-1)}+\mathcal{O}(2^{-m})\right)\nonumber\\
&=\frac{1}{(2k)!}(1+\mathcal{O}(2^{-m}))
\end{align} 
where we used $\left(\begin{array}{c}
k\\
j
\end{array}\right):=\frac{k!}{j!(k-j)!}$ to obtain line 2 and Pascals rule $\left(\begin{array}{c}
k+1\\
j
\end{array}\right)=\left(\begin{array}{c}
k\\
j
\end{array}\right)+\left(\begin{array}{c}
k\\
j-1
\end{array}\right)$ for line 3 and $\sum_{j=0}^k\left(\begin{array}{c}
k\\
j
\end{array}\right)=2^k$ for line 4.

Having shown (\ref{falloff-for-a}), we have also shown that the flow drives the initial value indeed to the fixed point as $n\rightarrow \infty$.\\

Studying the flow of $d_n(q)=b_n(q)+c_n(q)$ however is considerably more difficult. Although being described by the linear recursion relation $d_{n+1}(q)=4(1+a_n(q/2))d_n(q/2)$, not knowing the analytic form of the $a_n(q)$ entering in each step makes it analytically impossible to evaluate exactly whether $d_0=q^3/2$ flows indeed into $d^*(q)=q(\cosh(q)-1)$. Instead, we will present the numerical evidence, that it approaches the fixed point rather fast, see figure \ref{figure-flow}. We plot the functional dependence as a function of $q$ for different iterations in steps $n$. 
At fixed $n$ the deviation from the fixed point is bigger for higher values of $q$ since $a_n(q)$ is a polynomial of degree $2^n$. At fixed $q$ the deviation decreases as we increase $n$.
\begin{figure}[h]\label{figure-flow}
\begin{center}
\includegraphics[scale=0.635]{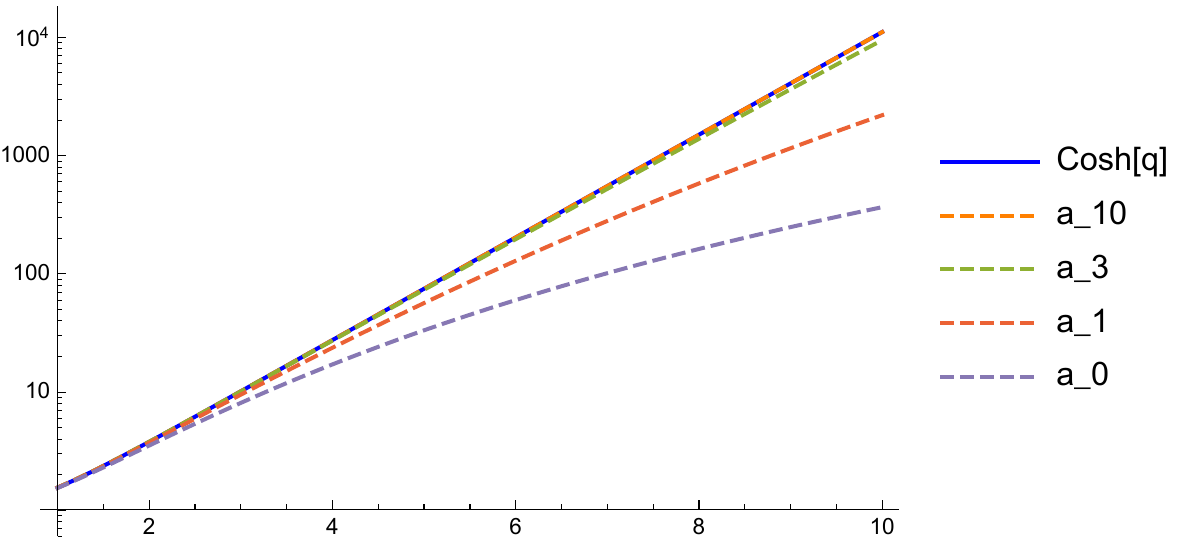}
\includegraphics[scale=0.635]{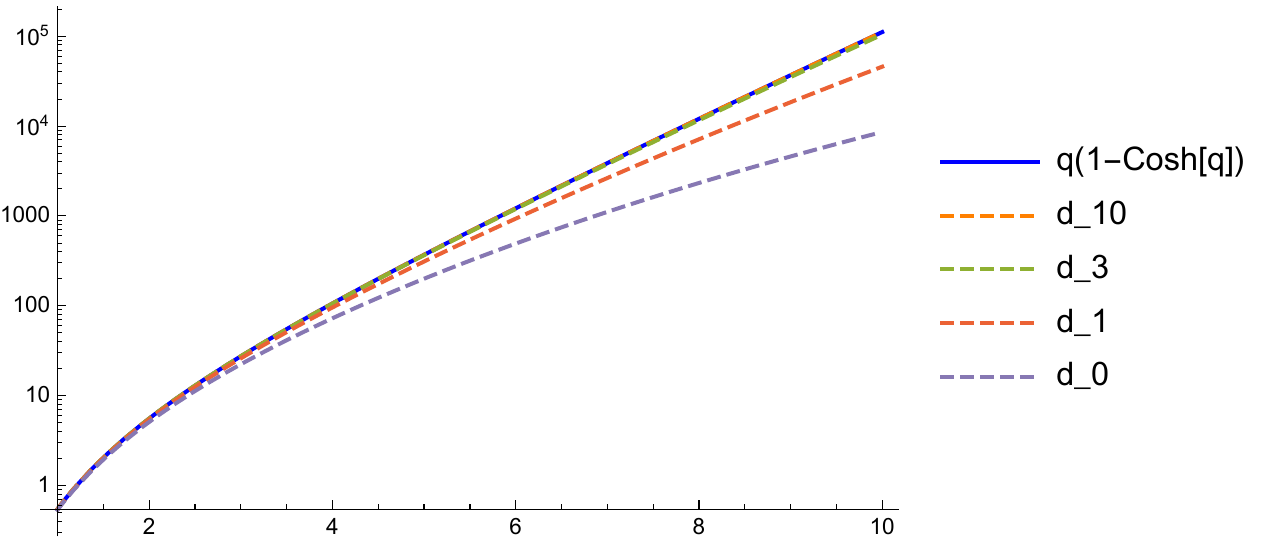}
\caption{\footnotesize  The flow of $a_n(q)$ (left) and $d_n(q)$ (right) for $q\in\{1,10\}$ in a logarithmic plot. We used dashed lines for the iterations $n=0,1,3,10$ in different colours and compare this to the corresponding fixed point functions ($a_*(q)={\rm ch}(q)$, $d_*(q)=q(1-{\rm ch}(q)$) as solid blue line. The fixed point is approached from below extremely fast in both cases, such that the blue line is almost indistinguishable from the orange dashed line for $n=10$ in the depicted interval.}
\end{center}
\end{figure}

\subsection{Stability properties with respect to initial conditions}
\label{stability}

While we have supplied analytical and numerical evidence that starting from the 
naive lattice Laplacian with only next neighbour contributions the 
flow indeed converges to the fixed point found, the question arises how stable 
the fixed point is under changing the initial discretisation of the covariance. In particular,
the fact that the flow studied in the previous section was parametrised by three functions only,
rests on the form of the initial discretisation. If we also consider initial discretisations
that involve next to next neighbour contributions, then a parametrisation by three functions
is no longer sufficient as we will see. The most general, translation invariant, 
symmetric form of the lattice Laplacian on a one dimensional lattice  consisting of $M$ points is given by (note the periodicity of the 
function)
\begin{align}
(\Delta_M f_M)(m)=\frac{1}{\epsilon_M^2}\sum_{k=0}^{\lfloor M/2\rfloor} \Delta_M(k) [f_M(m+k)+f_M(m-k)]
\end{align}
where the coefficients $\Delta_M(k)$ are subject to the constraint that for 
$f_M=E_M f,\; f\in C^\infty([0,R])$
the Taylor expansion up to second order in $\epsilon_M$ yields  $f^{\prime\prime}(m\epsilon_M)$.
We call this a physically allowed discretisation.  This gives the two constraints 
\begin{align}
\sum_{k=0}^{\lfloor M/2\rfloor}  \Delta_M(k)=0,\; \sum_{k=1}^{\lfloor M/2\rfloor} \Delta_M(k) k^2=1
\end{align}
leaving $\lfloor M/2\rfloor-2$ free parameters for the allowed discretisations. 
As an example, consider the next to next neighbour case, i.e. $\Delta_M(k)=0, k>2$ leaving one free parameter $\gamma$
\begin{align}\label{Laplacianfamily}
(\Delta_M^\gamma f_M)(m)=\frac{\epsilon^{-2}_M}{1+4\gamma}([f(m+1)+f(m-1)-2f(m)]+
\gamma[f(m+2)+f(m-2)-2f(m)])
\end{align}
The case $\gamma=0$ reproduces the naive next neighbour Laplacian, thus $\gamma\in \mathbb{R}$ labels 
its next to next neighbour type of perturbation.  

As an example, we consider a choice for $\gamma$ within the next to next neighbour discretisation 
class which makes $\Delta_M E_M f$ agree with $E_M \Delta f$ up to order $\epsilon_M^4$.
 The power expansion of $f(x\pm\epsilon),\;\epsilon\equiv \epsilon_M$ and $f(x\pm 2\epsilon)$ 
 results in the following linear system:
\begin{align}
\left(\begin{array}{c}
f(x+\epsilon)-f(x)\\
f(x+2\epsilon)-f(x)\\
f(x-\epsilon)-f(x)\\
f(x-2\epsilon)-f(x)
\end{array}\right)=\left(\begin{array}{cccc}
1 & 1/2 & 1/6 & 1/24\\
2 & 2 & 4/3 & 2/3\\
-1 & 1/2 & -1/6 & 1/24\\
-2 & 2 & -4/3 & 2/3
\end{array}\right)\left(\begin{array}{c}
\epsilon f'(x)\\
\epsilon^2 f''(x)\\
\epsilon^3 f^{(3)}(x)\\
\epsilon^4 f^{(4)}(x)
\end{array}\right)
\end{align}
If one inverts the appearing matrix, one can read of the contributions for $f''$ which are
\begin{align}
(\Delta f)(x) = f''(x) =\frac{1}{12\epsilon^2}\left(-f(x+2\epsilon)-f(x-2\epsilon)+16f(x+\epsilon)+16f(x-\epsilon)-30 f(x)\right)
\end{align}
This corresponds to the choice $\gamma=-\frac{1}{16}$.

Its eigenvalues in the Fourier basis are  (using $\cos(2x)=2\cos(x)^2-1$)
\begin{align}
(\Delta e^{ikn\;.}) (x) &= \frac{e^{iknx}}{12\epsilon^2}\left(-e^{ikn2}-e^{-ikn2}+16(e^{ikn}+e^{-ikn})-30\right)=\nonumber\\
&=-\frac{e^{iknx}}{6\epsilon^2}\left(\cos(2kn)-16\cos(kn)+15\right)=\nonumber\\
&=-\frac{1}{3\epsilon^2}e^{iknx}\left((\cos(kn)-4)^2-9\right)=:\hat{\Delta}(n)e^{iknx}
\end{align}
Accordingly, the initial covariance is now, described by (recall $q^2=(k_0^2+p^2)\epsilon_M^2$,
the appearing constants are irrelevant for what follows, see \cite{LLT1})
\begin{align} \label{initialdata}
\hat{C}^{(0)}_M(k_0,l)&
=R^{-1}\frac{\hbar\kappa}{2}\frac{3\epsilon^2_M}{(\cos(k_Ml)-4)^2-9+3\epsilon^2_M (p^2+k_0^2)}\nonumber\\
&=R^{-1}\frac{\hbar\kappa}{2}\frac{3\epsilon^2_M}{\cos(k_M l)^2-8\cos(k_Ml)+7+3 q^2}
\end{align}
Let $C^{(n)}_M=\frac{\hbar\kappa}{R}  c^{(n)}_M$.
The renormalisation flow (for $D=1$) is given by 
\begin{align}\label{RnMap}
\hat{c}^{(n+1)}_M(l,k_0)=\frac{1}{2}\left([1+\cos(k_{2M}l)]\hat{c}^{(n)}_{2M}(l,k_0)+[1-\cos(k_{2M}l)]\hat{c}^{(n)}_{2M}(l+M,k_0)\right)
\end{align}
We claim that this transformation leaves invariant the following functional form parametrised by 
six functions $a_n,...,f_n$ ($t=k_Ml$)
\begin{align}\label{generalNNcov}
\hat{C}^{(n)}_M(t)=\frac{\epsilon^2_M}{q^3}\frac{f_n(q)\cos(t)^2+e_n(q)cos(t)+d_n(q)}{c_n(q)\cos(t)^2-b_n(q)\cos(t)+a_n(q)}
\end{align}
The initial data can be read off from (\ref{initialdata})
\begin{align}\label{initsecondorder}
a_0=7+3q^2,\hspace{5pt} b_0=8,\hspace{5pt} d_0 = 3q^3,\hspace{5pt} e_0=0,\hspace{5pt}f_0=0, \hspace{5pt}c_0=1
\end{align}
After one renormalisation step, the denominator becomes the product
\begin{align}
2[ c_n &\cos(t)^2-b_n \cos(t) +a_n ] [ c_n \cos(t)^2+b_n\cos(t) +a_n ]=\nonumber\\
&=2\left(c^2_n\cos(t)^4+a_n^2+2a_nc_n\cos(t)^2-b_n^2\cos(t)^2\right)=\nonumber\\
&=2\left(c^2_n/4+c^2_n/4\cos(2t)^2+c^2_n/2\cos(2t)+a_nc_N\cos(2t)+a_nc_n-b_n^2\cos(2t)-b_n^2+a^2_n\right)=\nonumber\\
&=[c^2_n/2]\cos(2t)^2-[b^2_n-c^2_n-2a_nc_n]\cos(2t)+[c^2_n/2+2a_nc_n-b^2_n+2a^2_n]
\end{align}
Remembering that under the renormalisation we have $t\mapsto t/2,\; q\mapsto q/2$, we can read off the recursion relations
\begin{align}
\label{cflow}c_{n+1}(2q)&:= c_n(q)^2/2\\
\label{bflow}b_{n+1}(2q)&:= b^2_n(q)-c_n(q)\left(c_n(q)+2a_n(q)\right)\\
\label{aflow}a_{n+1}(2q)&:= c_n(q)\left(c_n(q)/2+2a_n(q)\right)-b^2_n(q)+2a_n^2(q)
\end{align}
For $n\to \infty$ we can immediately see that $c_n$ flows into the fixed point $c_*(q)=0$. Then the fixed point condition for (\ref{bflow}) becomes $b_*(2q)=b_*(q)^2$. This functional equation has the one parameter set of solutions 
$\alpha\mapsto b_*(q)=e^{\alpha q}$. Our initial condition (\ref{initsecondorder}) started with a function $b_0(q)$ that was even in $q$ and (\ref{bflow}) does not change this behaviour. Thus 
the only  choice is: $\alpha=0, b_*(q)= 1$.\\
Consequently, we find the fixed point condition for (\ref{aflow})
\begin{align}
a_*(2q)=2a_*(q)^2-1
\end{align}
already familiar from the next neighbour discretisation class and
which is solved by the functions $\cos$ and $\cosh$.\\
Looking now at the numerator from (\ref{RnMap})
\begin{align}
&2(1+\cos(t/2))\left[f_n \cos(t/2)^2+e_n \cos(t/2)+d_n\right]\left[c_n\cos(t/2)^2+b_n\cos(t/2)+a_n\right]+\nonumber\\
&+2(1-\cos(t/2))\left[f_n\cos(t/2)^2-e_n\cos(t/2)+d_n\right]\left[c_n\cos(t/2)^2-b_n\cos(t/2)+a_n\right]=\nonumber\\
&= [f_nc_n+f_nb_n+e_nc_n]\cos(2t)^2+\nonumber\\
&\;\;\;\;+2[f_nc_n+f_nb_n+e_nc_n+f_na_n-e_nb_n+d_nc_n+e_na_n+b_nd_n]\cos(2t)+\nonumber\\
&\;\;\;\;+2[f_nc_n/2+f_nb_n/2+e_nc_n/2+f_na_n+e_nb_n+d_nc_n+e_na_n+b_nd_n+2a_nd_n]
\end{align}
Hence the remaining recursion relations are
\begin{align}
\label{fflow}f_{n+1}(2q)&:=\left(f_nc_n+f_nb_n+e_nc_n\right)(q)\\
\label{eflow}e_{n+1}(2q)&:=2\left((f_n+e_n+d_n)c_n+(a_n+b_n)f_n+e_nb_n+d_nb_n+e_na_n\right)(q)\\
\label{dflow}d_{n+1}(2q)&:=2\left(\frac{1}{2}(2d_n+f_n+e_n)c_n+\frac{1}{2}(2a_n+b_n)f_n
+2a_nd_n+e_nb_n+b_nd_n+e_na_n\right)(q)
\end{align}
Plugging in the already known results  (i.e. $c_*=0$, $b_*= 1$ and $a_*\in \{\cosh(q), \cos(q)\}$) we find that the fixed point of $f_n$ must obey 
\begin{align}
f_*(2q)= f_*(q)
\end{align}
The only scale invariant function in one variable is a constant, i.e $f_*=K$. 
To see which value of $K$ is picked by the initial conditions it is sufficient to compute the 
flow at $q=0$. We notice that $d_0(0)=e_0(0)=f_0(0)=0$ and that 
 (\ref{fflow})-(\ref{dflow}) is a homogeneous system of equations of first order 
as far as the functions $d_n, e_n,f_n$ are concerned.
This means, by induction, that the values of $d_n,e_n,f_n$ at $q=0$ 
remain zero for the entire flow. It follows that: $K=0$.
The remaining fixed point conditions reduce then to those for the next neighbour class
discretisation. It follows that both the (unique) next neighbour class and the above 
choice from the next to next neighbour class have the same unique fixed point.   \\
\\
Concerning the convergence of the system towards the fixed point, the situation is more involved
than for the next neighbour class. While by similar methods $c_n(q)$ is explicitly computable as
\begin{align}
c_n(q)=2^{-\sum_{k=0}^n2^k}=2^{1-2^{(1+n)}}
\end{align}
it turns out that if we start with the initial values from (\ref{initsecondorder}) one finds that the flow of $a_n, b_n, d_n, e_n,$ $f_n$ 
for each coefficient of the respective power series diverges. Consider for instance
$a_n(0)$. As $c_n(q)$ approaches zero exponentially fast
 this means the for higher iterations we approach for $a_n(0)$ the recursive equation $a_{n+1}(0)=2a_n(0)^2-1$. 
Let $\delta_n=a_n(0)-1$ then $\delta_{n+1}=2\delta_n(2+\delta_n)$. This means that 
the error $\delta_n$ grows exponentially, i.e. $a_n(0)$ 
appears to be a {\it relevant coupling} in the 
terminology of statistical physics. For starting values $|a_0(0)|>1$ the sequence diverges.
For starting values $|a_0(0)|<1$ the sequence displays  chaotic behaviour 
and does not converge to the fixed point but there may be a subsequence that does.
Our chosen discretisation picks $a_0(0)=7$ so certainly $a_n(0)$ by itself does not converge.

Note however, that the convergence of the coefficient function sequences 
$a_n,..,f_n$ is only sufficient for the convergence of the covariance. Indeed, since 
the covariance is a homogeneous rational function of those six functions, that is, a fraction
with both numerator and denominator linear in those functions, after each renormalisation step
a common rescaling of those functions by any (non vanishing) other function such as 
a (non vanishing) constant leaves the covariance unaffected. It turns out that a common rescaling 
by $b_n(0)$ after each renormalisation step leads to modified sequences 
\begin{align}
a'_n(q):=\frac{a_n(q)}{b_n(0)},\;..\; f'_n(q):=\frac{f_n(q)}{b_n(0)}
\end{align} 
which now converge as the numerical evidence suggests. Even more, the convergence 
takes place independently of the value of $\gamma$ except for  $\gamma=\gamma_0=-\frac{1}{4}$ 
which plays a special role as the discretisation of the Laplacian blows up here.

\begin{figure}[H]
\begin{center}\label{figure-flow2}
\includegraphics[scale=0.67]{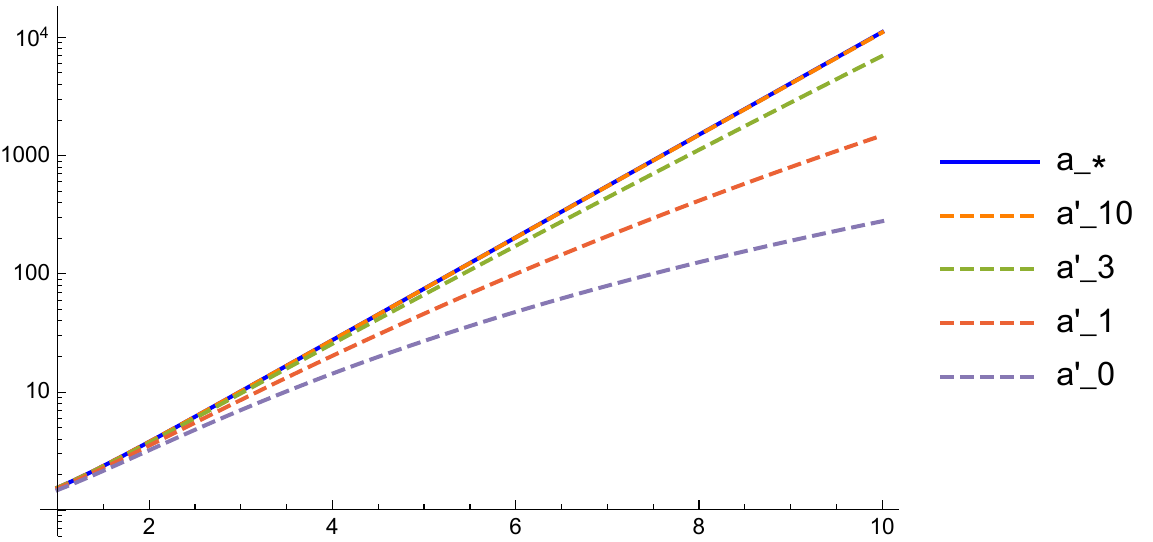}
\includegraphics[scale=0.67]{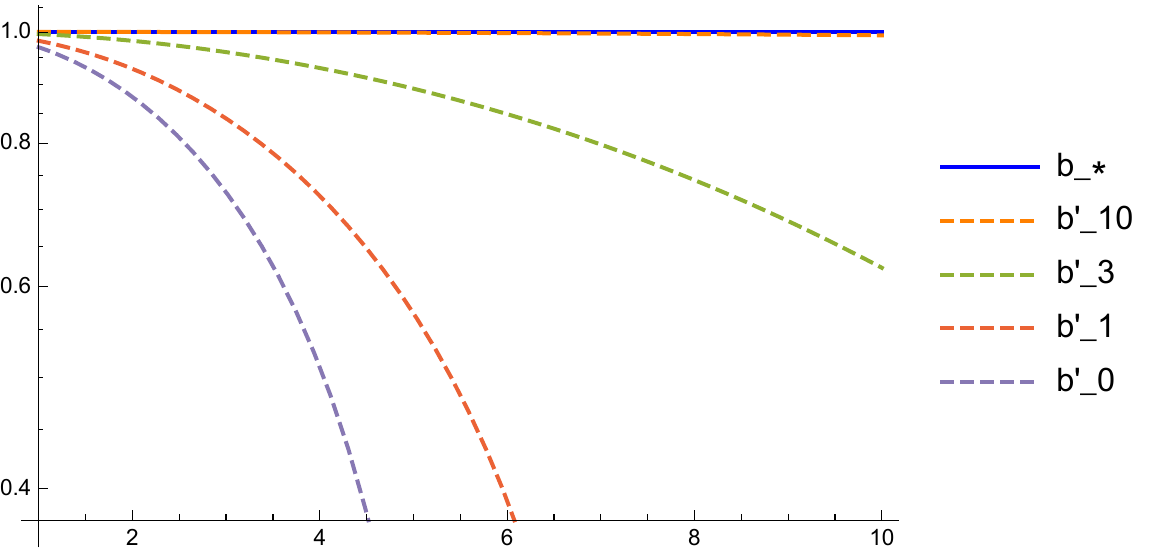}
\caption{\footnotesize The flow of the rescaled $a'_n(q)$ (left) and $b'_n(q)$ (right) for $q\in\{1,10\}$ in a logarithmic plot. We used dashed lines for the iterations $n=0,1,3,7,10$ in different colours and compare to this the corresponding fixed point functions ($a_*(q)={\rm ch}(q)$, $b_*(q)=1$) as a solid blue line. The fixed point is approached from below extremely fast in both cases, such that the blue line is almost indistinguishable from the orange dashed line for $n=10$ in the depicted interval. Note that we do not display $c'_n(q)$ which approaches zero exponentially fast.}
\end{center}
\end{figure}
\begin{figure}[H]
\begin{center}\label{figure-flow3}
\includegraphics[scale=0.67]{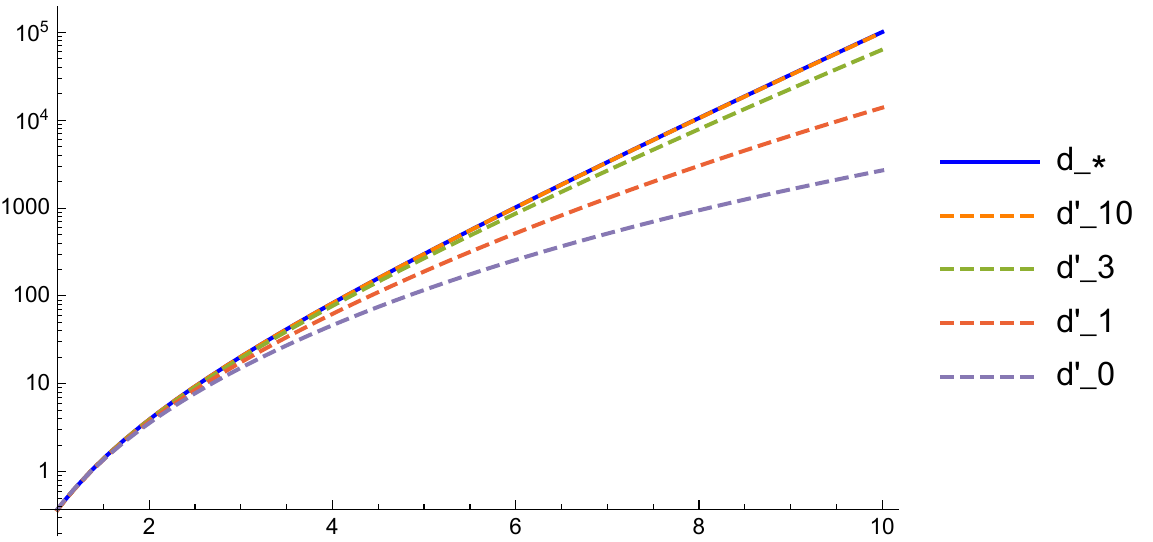}
\includegraphics[scale=0.67]{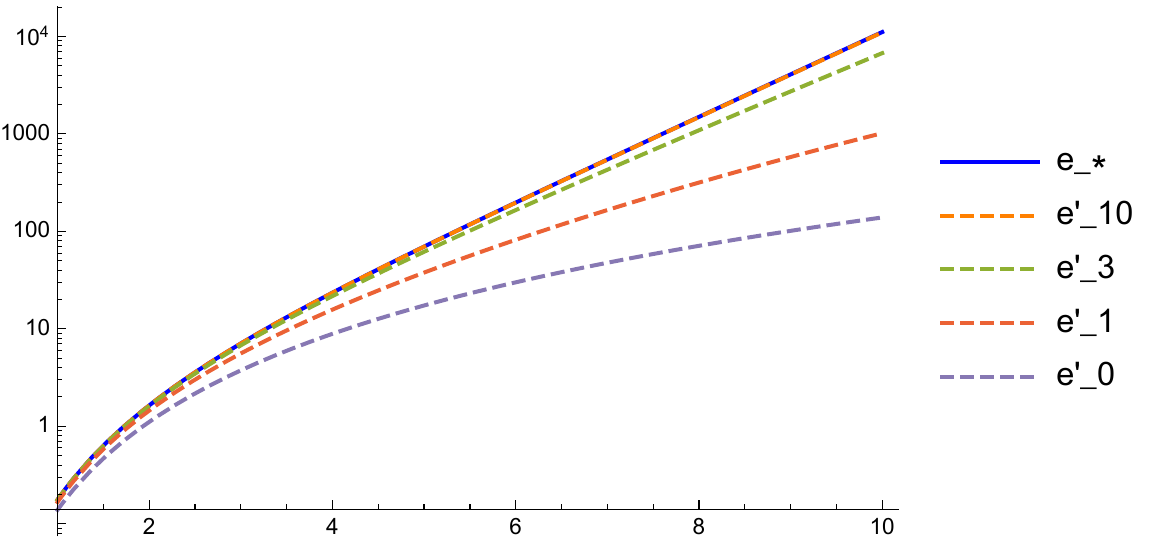}
\caption{\footnotesize The flow of the rescaled $d'_n(q)$ (left) and $e'_n(q)$ (right) for $q\in\{1,10\}$ in a logarithmic plot. We use dashed lines for the iterations $n=0,1,3,7,10$ in different colours and compare to this the corresponding fixed point functions ($d_*(q)=q{\rm ch}(q)-{\rm sh}(q)$, $e_*(q)={\rm sh}(q)-q$) as a solid blue line.  The fixed point is approached from below extremely fast in both cases, such that the blue line is almost indistinguishable from the orange dashed line for $n=10$ in the depicted interval. Note that we do not display  $f'_n(q)$ which approaches zero exponentially fast.}
\end{center}
\end{figure}
\begin{figure}[H]
\begin{center}\label{figure-flow4}
\includegraphics[scale=0.70]{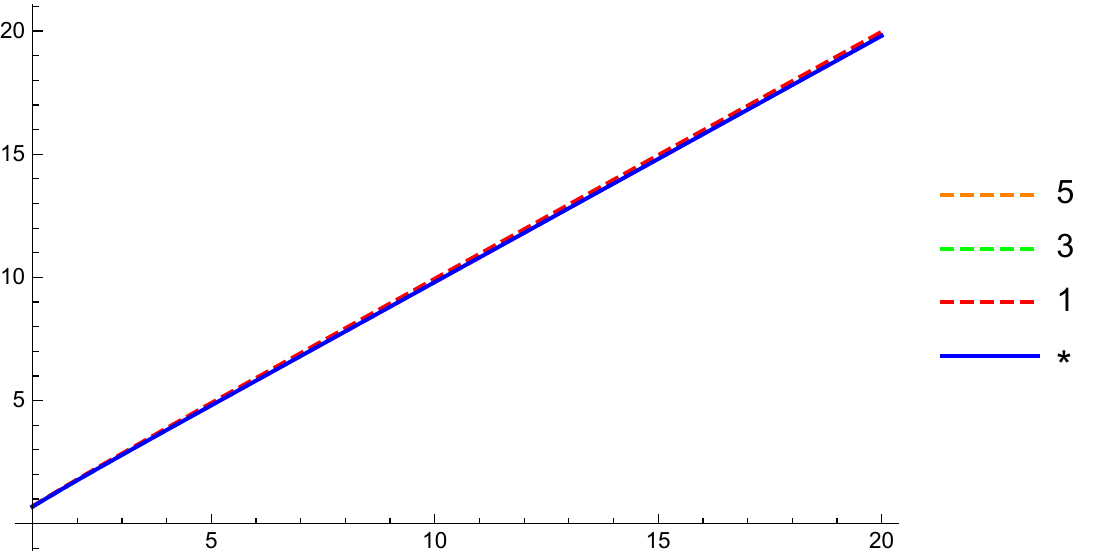}
\includegraphics[scale=0.70]{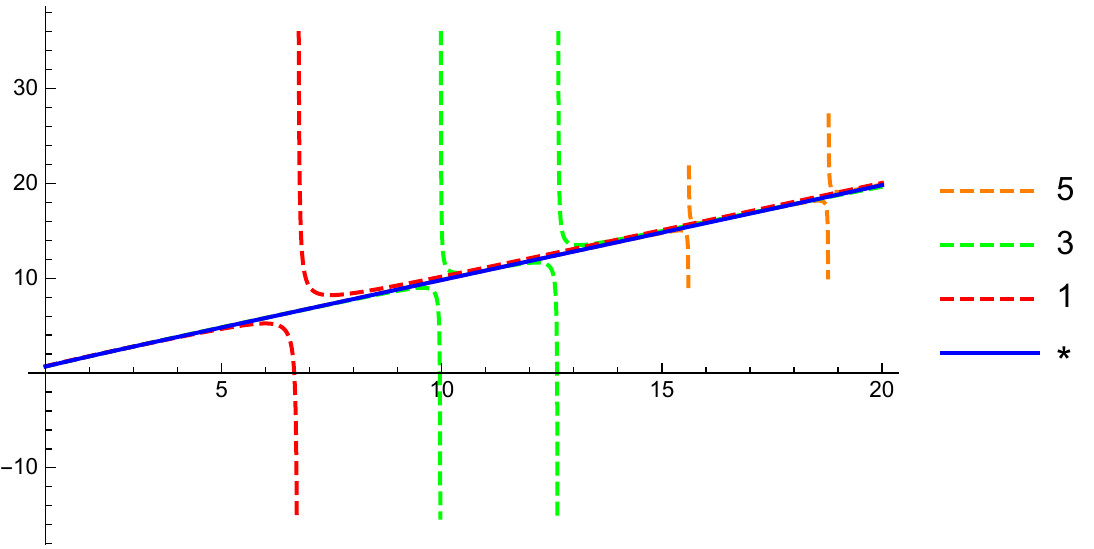}
\caption{\footnotesize The flow of the total covariance $\hat{C}^{(n)}(q,t=2\pi/10)$ 
as function of $q\in[1,20)$	for various iterations $n=1,3,5$ of the RG map. On the left the naive discretisation to start the flow with was $\gamma=1/3>\gamma_0$ and on the right it was $\gamma=-1/3<\gamma_0$. On the left, we find that the fixed point is approached very rapidly and on the right we see that, although the number of poles grows, they get shifted towards infinity for increasing $n$.}
\end{center}
\end{figure}
We plot both the individual functions at $\gamma=-1/16$ and the total covariance
at two values of $\gamma$ smaller and bigger than $\gamma_0$. The convergence 
of the covariance is faster 
for $\gamma>\gamma_0$ since for $\gamma<\gamma_0$ the denominator can become 
small. However the position of those minima moves to infinity as the flow proceeds. 
It is clear from this section how one would repeat the analysis eg. for the next to next to next  
neighbour class where one would have a two-parameter freedom. We leave this for future work. 

\subsection{Universality Properties}
\label{Uniqueness}

The flow - and correspondingly the final fixed point - of the measure family 
imposes the cylindrical consistency condition on the coarse graining maps, hence
only a certain subset of all possible coarse graining maps $I_{M\rightarrow 2M}$ can be considered. However, it will transpire that the {\it block-spin}-map considered so far is not unique. In this section we will consider others, e.g. the {\it deleting}-map, which is also cylindrically consistent. In order to bring our framework into contact with the literature, we will also consider this map in the path integral context and compare to already known results, see appendix \ref{standard}.\\
 \\
To review the notation, we consider the discretised Weyl element smeared with the test function $f_M$:
\begin{align}
w_M[f_M]=\exp\left(i\sum_{m\in\mathbb{Z}_M}\phi_M(m)f _M(m)\right)
\end{align}
which allows us to define the generating functional of the Hilbert space measure
 $\nu_M$ as:
\begin{align}
\nu_M(w_M[f_M])=\int d\nu_M(\phi_M) e^{i\phi_M[F_M]}
\end{align}
where in practice $d\nu_M(\phi^M)$ is a suitable weight function times $M$-copies of the Lebesque measure in $\phi_M(m)$.

The map $(I_{M\rightarrow 2M}f_M)(m)=f_M(\lfloor m/2\rfloor)$, $m\in\mathbb{Z}_{2M}$ which we had considered so far allows us to rewrite the {\it cylindrical consistency condition}
\begin{align}
\nu_M(w_M[f_M])=\nu_{2^nM}(w_{2^nM}[I_{M\rightarrow2^nM}f_M])
\end{align}
e.g. for $n=1$ as 
\begin{align}
\int d\nu_M(\phi_M)e^{i\phi_M[f_M]}=&\int d\nu_{2M}(\phi_{2M})e^{i\phi_{2M}[I_{M\rightarrow 2M}f_M]}=\nonumber\\
=&\int d\nu_{2M}(\phi_{2M})e^{i\sum_{m\in\mathbb{Z}_{2M}}f_M(\lfloor m/2\rfloor)\phi_{2M}(m)}\nonumber\\
=&\int d\mu_{2M}(\phi_{2M})e^{i\sum_{m'\in\mathbb{Z}_M}f_M(m')\left(\phi_{2M}(2m')+\phi_{2M}(2m'+1)\right)}
\end{align}
which shows that $I_{M\rightarrow 2M}$ indeed represents a block-spin-transformation in the 
usual sense.

However, the flow of measures certainly depends on the chosen block-spin-transformation
and thus also the fixed points could depend on it. The degree of independence of the 
choice of such a map $\tilde{I}_{M\rightarrow2M}$ is loosely referred to as {\it universality}.
For example, the  {\it deleting map} is defined by 
\begin{align}
\left(I^{Del}_{M\rightarrow2M}f_M\right)(m)=\begin{cases}
2^{\alpha} f_M(m/2) & \mbox{if } m\in2\mathbb{Z}_M\\
0 & \mbox{else}
\end{cases}
\end{align}
As one can check now, this map indeed passes the {\it the cylindrical consistency condition} for any $\alpha$
\begin{align}\label{conscond}
I^{Del}_{2^nM\rightarrow 2^{n+n'}M}\circ I^{Del}_{M\rightarrow 2^nM}=I^{Del}_{M\rightarrow2^{n+n'}M}
\end{align}
However, the only way to guarantee that this map is isometric is $\alpha=D/2$ because
\begin{align}
&\langle f_M,f'_M\rangle_M=\epsilon_M^D\sum_{m\in\mathbb{Z}^D_M}\bar{f}_M(m)f_M'(m)= \langle I^{Del}_{M\rightarrow2M}f_M,I^{Del}_{M\rightarrow2M}f'_M\rangle_M=\nonumber\\
&=\epsilon^D_{2M}\sum_{m\in2\mathbb{Z}_{2M}^D}2^{2\alpha}f_M(m/2)f'_M(m/2)=2^{2\alpha-D}\epsilon^D_M\sum_{m\in\mathbb{Z}_M^D}f_M(m)f'_M(m)
\end{align}
We refrain from constructing explicit evaluation and injection maps,  since they are irrelevant for determining the fixed point as well as taking the inductive limit. We will study it below and compare with the transformation used in \cite{LLT2}.
 
The set of coarse graining transformations satisfying cylindrical consistency is infinite
(e.g. one could use $I_{M\to p M}$ instead of $I_{M\to 2 M}$ where $p$ is any prime number).
Yet, these are indeed non-trivial conditions, and not all renormalisation-flows studied in the 
literature fulfil these conditions. E.g. in the literature 
it is standard to consider the {\it approximate blocking kernel} (see e.g. \cite{BW74,Has98,Has08,HN93} and reference therein)
\begin{align}\label{approxblockspin}
e^{i\phi_{2M}[I^{\kappa}_{M\rightarrow2M}F_M]}:=\int d\phi_Me^{-2\kappa[ \sum_{m\in\mathbb{Z}_M}(\phi_M(m)-\frac{1}{2}\sum_{m'\in\mathbb{Z}_{2M},
\lfloor m'/2\rfloor =m}\phi_{2M}(m')]^2}e^{i\phi_M[f_M]}
\end{align}
For $\kappa\rightarrow\infty$ the exponential tends to the  $\delta$-Dirac-distribution and reproduces the renormalisation map, $I_{M\rightarrow2M}=\lim_{\kappa\rightarrow\infty} I^{\kappa}_{M\rightarrow2M}$ which 
we have used so far and that satisfies the cylindrical consistency condition.

We review the renormalisation flow based on the approximate blocking transformation 
in appendix A. To see that our renormalisation flow $n\mapsto \nu^{(n)}_M$ is 
consistent with this standard flow defined in \ref{App2} in terms of action functionals 
we  write $d\nu^{(n+1)}_M(\phi_M)=d^M\phi_M e^{-\beta S^{(n+1)}_M(\phi_M)}$ with 
$S^{(n+1)}_M(\phi_M)$ being a function of $\phi_M$ (``action functional''), then:
\begin{align}
&\nu_M^{(n+1)}(w_M[f_M])=\int d^M\phi_M\left( e^{-\beta S^{(n+1)}_M(\phi_M)}\right) 
e^{i\phi_M[f_M]}=\int d\nu^{(n)}_{2M}(\phi_{2M})e^{i\phi_{2M}[I^{\kappa}_{M\rightarrow 2M}f_M]}
e^{i\phi_M[f_M]} 
\nonumber\\
& \overset{(\ref{approxblockspin})}{=}\int d^M\phi_M\left(
\int d\nu^{(n)}_{2M}(\phi_{2M})e^{-2\kappa[\sum_{m\in\mathbb{Z}}(\tilde{\phi}_M(m)-\frac{1}{2}\sum_{m'\in\mathbb{Z}_{2M},\lfloor m'/2\rfloor=m}\phi_{2M}(m')]^2}
\right)e^{i\phi_M[f_M]}
\end{align}

\subsubsection{Cylindrical Inconsistency of the Approximate Blocking Kernel}

We check whether (\ref{conscond}) holds for any $\infty>\kappa>0$. If true, then 
a necessary implication would be that
\begin{align}
e^{i\phi_{4M}[I^{\kappa}_{M\rightarrow 4M}f_M]} \overset{?}{=}e^{i\phi_{4M}[I^{\kappa}_{2M\rightarrow4M}\circ I^{\kappa}_{M\rightarrow 2M}f_M]}
\end{align}
which reads explicitly
\begin{align}\label{explconscond}
&\int d^{2M}\tilde{\phi}_{2M}e^{-2\kappa\sum_{m\in\mathbb{Z}_{2M}}[\tilde{\phi}_{2M}(m)-\frac{1}{2}\sum_{m'\in\mathbb{Z}_{4M},\lfloor m'/2\rfloor=m}\phi_{4M}(m')]^2}\times\nonumber\\
&\;\;\;\;\times\int d^M\phi'_M e^{-2\kappa\sum_{n\in\mathbb{Z}_M}[\phi'_M(m)-\frac{1}{2}\sum_{n\in\mathbb{Z}_{2M},\lfloor n'/2\rfloor =n}\tilde{\phi}_{2M}(n')]^2}\nonumber\\
&\overset{?}{=}\int d^M\phi'_M e^{-2\kappa \sum_{m\in\mathbb{Z}_M}[\phi'_M-\frac{1}{4}\sum_{m'\in\mathbb{Z}_{4M},\lfloor m'/4\rfloor =m}\phi_{4M}(m')]^2}e^{i\phi'_M[f_M]}
\end{align}
We evaluate the Gaussians on the left hand side:
\begin{align}
\int d^{2M}\tilde{\phi}_{2M}&\exp({-2\kappa \sum_{m\in\mathbb{Z}_{2M}}\tilde{\phi}_{2M}(m)^2-\kappa/2\sum_{m\in\mathbb{Z}_{2M}}\tilde{\phi}_{2M}(m)^2})\times\nonumber\\
&\times \exp({-\kappa/2\sum_{n\in\mathbb{Z}_M}\tilde{\phi}_{2M}(2n)\tilde{\phi}_{2M}(2n+1)})\exp({-2\kappa\sum_{m\in\mathbb{Z}_{2M}}\tilde{\phi}_{2M}(m)A(m)})
\end{align}
where we defined $A(m):=\sum_{m'\in\mathbb{Z}_{4M},\lfloor m'/2 \rfloor =m}\phi_{4M}(m')+\phi'_M(\lfloor m/2\rfloor)$.\\
We perform the integrals over $\tilde{\phi}_{2M}(2n+1),\;n\in \mathbb{Z}_M$ first and then 
perform the remaining integral over $\tilde{\phi}_{2M}(2n),\;n\in \mathbb{Z}_M$, denoted by 
$d^M\tilde{\phi}_{2M}$, resulting in
\begin{align}
&\int d^M\tilde{\phi}_{2M}\sqrt{\frac{2\pi}{5\kappa}}^Me^{\kappa/10\sum_{n\in\mathbb{Z}_M}(\frac{1}{2}\tilde{\phi}_{2M}(2n)+2A(2n+1))^2}\times\nonumber\\
&\;\;\;\;\times e^{-5\kappa/2\sum_{n\in\mathbb{Z}_M]}\tilde{\phi}_{2M}(2n)^2}e^{-2\kappa\sum_{n\in\mathbb{Z}_M}\tilde{\phi}_{2M}(2n)A(2n)}=\nonumber\\
&=\sqrt{\frac{2\pi}{5\kappa}}^Me^{2\kappa/5 \sum_{n\in\mathbb{Z}_M}A(2n+1)^2 }
\int d^M\tilde{\phi}_{2M}\times\nonumber\\
&\;\;\;\;e^{\sum_{n\in\mathbb{Z}_M}\tilde{\phi}_{2M}(2n)^2(1/40-5/2)\kappa}
e^{-\sum_{n\in\mathbb{Z}_M}\tilde{\phi}_{2M}(2n)\left(2\kappa A(2n)-\kappa/5 A(2n+1)\right)}
=\nonumber\\
&=\sqrt{\frac{2\pi}{5\kappa}}^M\sqrt{\frac{40\pi}{99\kappa}}^M e^{\frac{2\kappa}{5}\sum_{n\in\mathbb{Z}_M}A(2n+1)^2}e^{\frac{10}{99}\kappa\sum_{n\in\mathbb{Z}_M}(2A(2n)-1/5A(2n+1))^2}
\end{align}
It is transparent, that e.g. the coefficients  of the $\phi'_M(n)^2,\;n\in\mathbb{Z}_M $
appearing in the exponent of the last line above  do not sum up to $-2\kappa$. It follows that $I^{\kappa}_{M\rightarrow 2M}$ does not fulfil the cylindrical 
consistency condition for any finite $\kappa$ and we exclude it from the list of acceptable 
blocking kernels.

\subsubsection{Continuum theory for different blocking-kernels}

Both the deleting kernel and the kernel we used so far are cylindrically consistent.
How do their flows compare to each other?\\
\\
To answer this question, we investigate the flow of
\begin{align}
C^{Del,(n+1)}_M = (1_{L_T} \otimes I^{Del}_{M\rightarrow2M})^{\dagger} C^{Del,(n)}_{2M} (1_{L_T} \otimes I^{Del}_{M\rightarrow2M})
\end{align}
which can be computed using the same methods as in \cite{LLT2}:
\begin{align}
&\langle F_M, C^{Del,(n+1)}_M F_M\rangle_M =\nonumber\\
&= \epsilon^{2D}_{2M}\sum_{m_1,m_2\in\mathbb{Z}^D_{2M}}\int ds \int ds' \times\nonumber\\
&\;\;\;\;\times (1_{L_T} \otimes I^{Del}_{M\rightarrow2M}F_M)(s,m_1)(1_{L_T} \otimes I^{Del}_{M\rightarrow2M}F_M)(s',m_2) C^{Del,(n)}_{2M}((s,m_1),(s',m_2))=\nonumber\\
&=\frac{2^{2\alpha}}{2^{2D}}\epsilon^{2D}_M \int ds \int ds' \sum_{m_1,m_2\in\mathbb{Z}^D_M}F_M(s,m_1)F_M(s',m_2)C^{Del,(n)}_{2M}(s,m_1),(s',m_2))
\end{align}
Which tells us that the flow of the covariance is given by
\begin{align}\label{DeletingFlow}
C^{Del,(n+1)}_M((s,m_1),(s',m_2))=2^{2(\alpha-D)}C^{Del,(n)}_{2M}((s,m_1),(s',m_2))
\end{align}
and consequently, also for their discrete Fourier transforms. We find this to drive the $D=1$ starting covariance
\begin{align}
\hat{C}^{(0)}_M(k_0,l)=R^{-1}\frac{\hbar\kappa}{2}\frac{1}{\epsilon^{-2}_M(1-\cos(k_M l))+k_0^2+p^2}
\end{align}
to zero or infinity unless $\alpha=D$. This demonstrates two things: First, the isometry of the 
coarse graining map is not a necessary condition in order to define a suitable flow.
On the other hand, by far not every map defines a meaningful flow.

Picking $\alpha=D$  we compare the continuum limits $M\rightarrow\infty$ of both 
fixed point covariances computed by the block spin and deleting kernel respectively
\begin{align}
\lim_{M\rightarrow\infty} C^*_M(k_0,l)=\frac{\hbar\kappa}{4R}\frac{1}{p^2+k_0^2+(2\pi l)^2}
=\lim_{M\rightarrow\infty} C^{Del,*}_M(k_0,l)
\end{align}
Thus the two continuum theories they define are identical when $\alpha=D$.
Note that, trivially, the cylindrical projections of the same continuum covariance with respect 
to two different projections corresponding to different blocking kernels are of course
different. 

\subsection{Perfect Lattice Laplacian}
\label{s4}

The adjective ``perfect'' is used in renormalisation theory in order to characterise quantities 
at finite resolution of the fixed point theory. For instance, the family of fixed point 
covariances labelled by the finite resolution (lattice) parameter $M$ in the free field theory case defines an effective 
covariance which can be interpreted as the result of integrating the spacetime Weyl element 
against the exponential of an effective action at the given resolution. This action is called ``perfect action'', actually a whole family thereof.
In this section we investigate the family of ``perfect Laplacians'' which can be extracted from the
family of fixed point covariances and study the decay behaviour of the contribution of lattice 
points in $r$-th neighbour relation to a given lattice point.

To avoid confusion, in the literature the term ``perfect lattice Laplacian'' mostly refers to the 
Euclidian d'Alembert operator, i.e. the operator $\Box=\partial_t^2+\Delta$ involving time 
where $\Delta$ is the spatial Laplacian. In our case, we are more interested in the 
``perfect spatial lattice Laplacian'' which refers to $\Delta$. These two quantities are 
defined in terms of the finite resolution operators given by the fixed point theory. We have direct 
access to the fixed point family of spacetime covariances 
\begin{align}
M\mapsto C^\ast_M:=(1_{L_2}\otimes I_M)^\dagger (p^2-\Box)^{-1} 
(1_{L_2}\otimes I_M)
\end{align}
of our renormalisation flow whose 
Fourier transforms $\hat{C}^\ast_M(k_0,l),\; l\in \mathbb{Z}_M$ were explicitly computed. 
We can now define the perfect Euclidian 
d'Alembertian as $-\Box^\ast_M:=(C^\ast_M)^{-1}-p^2$. Recall the continuum covariance (dropping all prefactors, $\epsilon_M=\frac{R}{M}$, $k_M=\frac{2\pi}{M}$)
\begin{align}
C=\frac{1}{2}(-\partial_t^2-\Delta+p^2)^{-1}
\end{align}
The initial datum for the RG-flow was defined in terms of the naively discretised Laplacian, i.e. $(\Delta^{(0)}_Mf_m)(m)=(f_M(m+1)+f_M(m-1)-2f_M(m))/\epsilon_M^2$, with covariance 
\begin{align} \label{transferofFT}
C^{(0)}_M=\frac{1}{2} (-\partial_t^2-\Delta^{(0)}_M+p^2)^{-1}
\end{align}
Its flow in Fourier space gave the fixed point
\begin{align}
\frac{2}{\epsilon^2_M}\hat{C}^*_M(k_0,l)=\frac{1}{q^3}\frac{\left(\cosh(q)q-\sinh(q)+(\sinh(q)-q)\cos(t)\right)}{\cosh(q)-\cos(t)}
\end{align}
with $t=k_M,q^2=(k_0^2+p^2)\epsilon_M^2$. The Fourier transform of the 
perfect d'Alembertian family 
$M\mapsto \Box^*_M$ is given by the inverse of (\ref{transferofFT}):
\begin{align}
-\hat{\Box}_M^*+p^2:=\epsilon^{-2}_M \frac{q^3[\text{ch}(q)-\cos(t)]}{\text{ch}(q)q-\text{sh}(q)+(\text{sh}(q)-q)\cos(t)}
\end{align}
The partially discrete kernel $(\Box^*_M F_M)(s,m)=:\int ds'  \sum_{m'\in\mathbb{Z}_M}\Box_M^*(s-s',m-m')F_M(s',m')$ reads explicitly
\begin{align} \label{perfectkernel}
\Box^*_M(s,r)=\int \frac{dk_0}{2\pi}\;\frac{1}{M}\sum_{l\in\mathbb{Z}_M}e^{ik_Mlr+isk_0}
[p^2-\epsilon^{-2}_M \frac{q^3[\text{ch}(q)-\cos(t)]}{\text{ch}(q)q-\text{sh}(q)+(\text{sh}(q)-q)\cos(t)}]
\end{align}

We want to find out whether  
that $\Box^\ast_M(s,r)$ decays exponentially fast with the spatial neighbour parameter $r$.
To do this, we define the forward and backward lattice shifts as follows
\begin{align}
(\delta^{+k} f)(m)=f(m+k),\hspace{20pt}(\delta^{-k} f)(m)=f(m-k)
\end{align}
with $k=-\lfloor M/2\rfloor,...,\lfloor M/2\rfloor$ which implies $(\delta^+)^n=\delta^{+n}$ and $\delta^+\delta^- =\delta^-\delta^+$. Note that $\delta^{\pm k+\alpha M}=\delta^{\pm k}$ for all $\alpha\in\mathbb{Z}$. 

Now $\cos(t)$ is an eigenvalue of $\mathfrak{d}:=[\delta^{+} +\delta^{-}]/2$ in Fourier space 
\begin{align}\label{EigenvalueOfMathDelta}
(\mathfrak{d} e^{it \; .})(m)=\frac{1}{2}(e^{itm+it}+e^{itm-it})=cos(t)e^{itm}
\end{align}
so that 
 \begin{align} \label{perfectkernel1}
\Box^*_M(s,r)=\int \frac{dk_0}{2\pi}\;e^{isk_0}\;
([p^2-\epsilon^{-2}_M \frac{q^3[\text{ch}(q)-\mathfrak{d}]}{\text{ch}(q)q-\text{sh}(q)+(\text{sh}(q)-q)\mathfrak{d}}] \cdot \;^K\delta_0)(r)
\end{align}
where $r\mapsto \;^K\delta_{0,r}\equiv \;^K\delta(0,r)$ is the Kronecker $\delta$ supported at $0$ and the operator $\mathfrak{d}$ acts on the variable $r$ in this formula. Similarly, we may introduce the operator  $Q^2:=\epsilon_M^2 (p^2-\partial_s^2)$ and the function $A(Q):=(\text{sh}(Q)-Q)/(Q\text{ch}(Q)-\text{sh}(Q))$. Then 
\begin{align} \label{perfectkernel2}
\epsilon^2_M \Delta^*_M (s,r):= \epsilon^2_M (\Box^*_M-\partial_s^2)(s,r)=([Q^2-\frac{Q^3}{Q\text{ch}(Q)-\text{sh}(Q)}\frac{\text{ch}(Q)-\mathfrak{d}}{1+A(Q) \mathfrak{d}}] \cdot 
\delta_0\otimes \;^K\delta_0)(s,r)
\end{align} 
where $\delta_0$ is the Dirac $\delta$ distribution for the temporal degree of freedom.\\
The first term in (\ref{perfectkernel2}) gives a contribution on $r=0$ only. Hence, to study the decay behaviour for spatial directions, we focus on the second term: By integrating respectively summing (\ref{perfectkernel1}) against time-independent functions $f(s,r)=f_M(r),f'(s',r')=f'_M(r')$ we obtain $\delta_{k_0,0}$, in other words $Q^2=q^2_0:=p^2\epsilon_M^2$ and
\begin{align}
\langle f',\Box^*_M f\rangle=\langle f'_M, \Delta^*_M f_M\rangle_M
\end{align}
The idea is now to expand its denominator into a geometric series with respect to the operator $\mathfrak{d}$ and to extract the coefficients of $\delta^{\pm k}$.   To expand it into a Neumann-series, we must check for convergence of the series. This will be guaranteed if $||A(q_0) \mathfrak{d}|| \leq 1$ in the operator norm $||.||$.\\
First, note that $A(q_0)\leq \frac{1}{2}$ for all $q_0\geq 0$, because
\begin{align}
2 (\text{sh}(q_0)-q_0) \leq q_0\text{ch}(q_0)-\text{sh}(q_0)\; &\Leftrightarrow\nonumber\\
3 \text{sh}(q_0) \leq q_0 \text{ch}(q_0)+2q_0\;&\Leftrightarrow\nonumber\\
\sum_k \frac{3}{(2n+1)!}q_0^{2n+1}\leq  \sum_k \frac{1}{(2n)!}q_0^{2n+1}+2q_0\;&
\end{align}
which can be checked by comparing all powers of $q_0$ separately.\\
Since $\delta^{\pm}$ are norm preserving, we use the Cauchy-Schwarz inequality to see that $||\mathfrak{d}||\leq 1$. Thus, on the functions of independent time support
\begin{align}
||A(q_0)\mathfrak{d}|| = |A(q_0)| \cdot ||\mathfrak{d}||\leq 1/2
\end{align}
and we can expand (\ref{perfectkernel2}) into a geometric series.\\
This gives
\begin{align}\label{SplittedSeries}
&\frac{\text{ch}(q_0)-\mathfrak{d}}{1+A(q_0)\mathfrak{d}}=(\text{ch}(q_0)-\mathfrak{d})\sum_{N=0}^\infty (-\mathfrak{d}A(q_0))^N=\\
&\;=\sum_{N=0}^\infty (-A(q_0))^N\left(\text{ch}(q_0)2^{-N}[\delta^++\delta^-]^N-2^{-N-1}[\delta^++\delta^-]^{N+1}\right)=\nonumber\\
&\;=\sum_{N=0}^\infty (-A(q_0)/2)^{N}\left(\text{ch}(q_0)\sum_{k=0}^N\left(\begin{array}{c}
N\\
k
\end{array}\right)\delta^{+k}\delta^{-(N-k)}-\frac{1}{2}\sum_{k=0}^{N+1}\left(\begin{array}{c}
N+1\\
k
\end{array}\right)\delta^{+k}\delta^{-(N+1-k)}\right)=\nonumber\\
&=\sum_{r\in\mathbb{Z}} \delta^{+2r}\left(\text{ch}(q_0)\sum_{n=0}^\infty\left(\begin{array}{c}
2n\\
n+r
\end{array}\right)(-A(q_0)/2)^{2n}-\frac{1}{2}\sum_{n=1}^\infty \left(\begin{array}{c}
2n\\
n+r
\end{array}\right)(-A(q_0)/2)^{2n-1}\right)+\nonumber\\
&\;\;\;+\sum_{r=1}^\infty\delta^{+2r-1}\left(\text{ch}(q_0)\sum_{n=0}^\infty\left(
\begin{array}{c}
2n+1\\
n+r
\end{array}\right)(-A(q_0)/2)^{2n+1}-\frac{1}{2}\sum_{n=0}^\infty \left(\begin{array}{c}
2n+1\\
n+r
\end{array}\right)(-A(q_0)/2)^{2n}\right)+\nonumber\\
&\;\;\;+\sum_{r=1}^\infty \delta^{-2r+1}\left(\text{ch}(q_0)\sum_{n=0}^\infty \left(\begin{array}{c}
2n+1\\
n-r+1
\end{array}\right) (-A(q_0)/2)^{2n+1}-\frac{1}{2}\sum_{n=0}^\infty\left(\begin{array}{c}
2n+1\\
n-r+1
\end{array}\right)(-A(q_0)/2)^{2n}\right) \nonumber
\end{align}
where have chosen $2r:=k-(N-k)\Rightarrow N=:2n$ and $2r:=k-(N+1-k)\Rightarrow N=:2n-1$ respectively for the even powers of $\delta^{\pm}$ and similar for the odd contributions. During this procedure, we used Fubini's theorem to exchange the summation order of $r,n$.\\
Indeed, for $A(q)\leq 1/2$ each sum over $n$ converges separately:
\begin{align}
\sum_{n=0}^\infty\left(\begin{array}{c}
2n\\
n+r
\end{array}\right)\cdot(A(q)/2)^{2n}\leq  \sum_{n=0}^\infty \frac{(2n)^{n+r}}{(n+r)!}\cdot 4^{-2n} \leq 
\sum_{n=0}^\infty \frac{4^{-2n}}{e}\left(\frac{2n\; e}{(n+r)}\right)^{n+r}=\nonumber\\
=\sum_{n=0}^\infty 2^r e^{r-1}\left(\frac{1}{1+r/n}\right)^{n+r} \left(\frac{e}{8}\right)^n \leq \sum_{n=0}^\infty 2^r e^{r-1} \left(\frac{e}{8}\right)^n = \frac{2^{r+3} e^{r-1}}{8-e}
\end{align}
where we have used a standard approximation for the factorial, i.e. $(n/e)^n e \leq n!$, and summed a geometric series. Thus, the inner sums over $n$ in (\ref{SplittedSeries}) are finite. The convergence and
\begin{align}
\sum_{n=0}^\infty\left(\begin{array}{c}
2n\\
n+r
\end{array}\right)z^{2n}=\sum_{n=0}^\infty\left(\begin{array}{c}
2n\\
n-r
\end{array}\right)z^{2n}=\sum_{k=0}^\infty\left(\begin{array}{c}
2k+2r\\
k
\end{array}\right) z^{2k+2r}
\end{align}
allow to identify the series with a {\it generalised binomial series} $\mathcal{B}_t(z)$. These kinds of sums were introduced by Lambert in 1758 \cite{Lam1758} and he showed later that its powers $r\in\mathbb{Z}$ obey the following property \cite{Lam1770}
\begin{align}
\mathcal{B}_t(z)^r=\sum_{k= 0}^\infty\left(\begin{array}{c}
tk+r\\
k
\end{array}\right)\frac{r}{tk+r}z^k
\end{align}
$\forall t\in\mathbb{Z}$ and $z\in\mathbb{R}$ such that the series converges. A modern proof of this statement can be found in \cite{Bala15}. Further, we quote the following identities from \cite{Grah94}
\begin{align}
\frac{\mathcal{B}_2(z)^r}{\sqrt{1-4z}}=\sum_{k= 0}^\infty \left(\begin{array}{c}
2k+r\\
k
\end{array}\right) z^k,\hspace{30pt} \mathcal{B}_2(z)=\frac{1-\sqrt{1-4z}}{2z}
\end{align}
Using this, we can compute the series explicitly: Let $s\in \{0,1\}$
\begin{align}
&\sum_{n=0}^\infty\left(\begin{array}{c}
2n+s\\
n+r
\end{array}\right) (-A(q_0)/2)^{2n}=\sum_{k=0}^\infty\left(\begin{array}{c}
2k+2r-s\\
k
\end{array}\right)(A(q_0)/2)^{2k}(A(q_0)/2)^{2r-2s}=\nonumber\\
&\;\;=\frac{(A(q_0)/2)^{2r-2s}}{\sqrt{1-4(A(q_0)/2)^2}}\mathcal{B}_2(A(q_0)^2/4)^{2r-s}=\frac{(A(q_0)/2)^{2r-2s}}{\sqrt{1-A(q_0)^2}}\left(\frac{2}{A(q_0)^2}\right)^{2r-s}(1-\sqrt{1-A(q_0)^2})^{2r-s}\nonumber\\
&\;\;=\frac{2^s(1-\sqrt{1-A(q_0)^2})^{-s}}{\sqrt{1-A(q_0)^2}}\left(\frac{1-\sqrt{1-A(q_0)^2}}{A(q_0)}
\right)^{2r}\sim \exp\left( r\cdot 2\log[1/A(q_0)-\sqrt{1/A(q_0)^2-1}]\right)\nonumber\\
&\;\;=:e^{r \Theta(q_0)}
\end{align}
For finite $q_0$, we have $0 < A(q_0)\leq 1/2$ and the logarithm is always well-defined and negative, since it holds that $1-\sqrt{1-A(q_0)^2}\leq A(q_0)$.\\
Thus, the perfect spatial lattice Laplacian is on the subspace of functions of independent time support explicitly given by
\begin{align}\label{perfLaplaceResult}
\epsilon^2_M \Delta^*_M=(q_0^2+&\frac{q_0^3A(q_0)^{-1}}{q_0 \text{ch}(q_0)-\text{sh}(q_0)}) \;{\rm id}-\frac{q_0^3\sqrt{1-A(q_0)^2}^{-1}}{q_0 \text{ch}(q_0)-\text{sh}(q_0)}\left(\text{ch}(q_0)A(q_0)+1\right)\times\\
&\;\;\times\left( \sum_{r\in\mathbb{Z}} \delta^{+2r}\frac{e^{|r|\Theta(q_0)}}{A(q_0)}-
\sum_{r=1}^\infty (\delta^{+(2r-1)}+\delta^{-(2r-1)})\frac{e^{|r|\Theta(q_0)}}{1-\sqrt{1-A(q_0)^2}}\right)\nonumber
\end{align}
\\
Lastly, we must account for the periodic boundary conditions. Remembering that the lattice identifies the points $r$ and $r+\alpha M$ with $\alpha \in \mathbb{Z}$, we add all corresponding contributions together. For the even powers of the shift operator:
\begin{align}
&\sum_{r\in\mathbb{Z}}\delta^{2r}e^{|r|\Theta(q_0)}=\sum_{r=-\lfloor M/4\rfloor}^{\lfloor M/4\rfloor}\delta^{2r}e^{|r|\Theta (q_0)}\sum_{\alpha\in\mathbb{Z}} e^{M |\alpha | \Theta(q_0)}=\nonumber\\
&\;\;\;\;\;=\sum_{r=-\lfloor M/4\rfloor }^{\lfloor M/4\rfloor }\delta^{2r} e^{|r|\Theta (q_0)}[\frac{2}{1-(\frac{1}{A(q_0)}-\sqrt{\frac{1}{A(q_0)^2}-1})^{2M}}-1]\nonumber
\end{align}
and the same geometric sum appears for the odd powers. For big lattices, i.e. $M>>1$, we see that due to $0<A(q_0)\leq 1/2$ the term in the brackets $[...]$ approaches $1$ very fast.\\
In total, we conclude that the perfect spatial Laplacian decays exponentially with $r$ and has a damping factor of $\Theta(q_0=p\epsilon_M^2)$. So, although it features non-local contributions, these are highly suppressed.\\

\section{Summary and Outlook}
\label{s5}

This is the third paper in a series of four in which we analyse a possible Hamiltonian renormalisation scheme motivated by the canonical approach to quantum gravity \cite{Rov04,AL04,Thi07}. As is well known, in quantum gravity the interaction is not even polynomial in the fields and thus tremendous additional complications arise in the path integral approach as compared to matter quantum fields on Minkowski space. This appears to make the canonical approach more tractable, although the same steps (regularisation, renormalisation) have to be performed. See the first paper \cite{LLT1} in this series for more details.

In this paper we examined the properties of the Hamiltonian renormalisation flow 
of the free two-dimensional Klein-Gordon field computed in \cite{LLT2} and thereby 
could make contact to the standard notions of renormalisation theory such as 
universality, stability, criticality, (ir)relevance and perfectness. This confirms that
the Hamiltonian renormalisation scheme delivers sensible and expected results 
although the time coordinate plays a special role as compared to path integral 
renormalisation, at least in the free field theory case.

Of course our real interest lies in the case of interacting quantum field theories, 
in particular those, which are not quantised using linear field variables but rather
variables that are common in gauge field theories such as holonomy variables.
The corresponding Hamiltonian
family is then in close analogy to Wilson kind of actions or Hamiltonians for non-Abelian gauge theories and 
represents additional challenges during the renormalisation process. We will come back to this issue in a future publication.

\appendix
\section{Standard Renormalisation}
\label{standard}

In this appendix we review path integral renormalisation as done in the mainstream 
of the literature based on seminal work by Wilson, Bell and Hasenfratz et. al. using the 
example of the massless 2-dimensional Klein-Gordon field and the averaging blocking kernel \cite{BW74, Has98,Has08,HN93}.
We apply their methods for the first time also to the deleting blocking kernel in the last section.  \\
\\
We start with the Euclidian action for a free massless scalar particle in $d=1+1$:
\begin{align*}
S:= \frac{\beta}{2}\int_{\mathbb{R}^{D+1}}dtd^Dx[\frac{1}{c}\dot{\Phi}^2-c\Phi\omega^2\Phi]=:\frac{\beta}{2}\int d^dx \Phi \Omega^2 \Phi
\end{align*}
and a discretisation thereof ($\epsilon_M:=1/M$):
\begin{align}
S_M(\phi_M)=: \frac{\beta\epsilon_M^2}{2}\sum_{n\in\mathbb{Z}^2_M}\sum_{m\in\mathbb{Z}^2_M}\Phi_M(m)\Omega_M^2(m,n)_M\Phi_M(n)
\end{align}
where $M$ is the UV cut-off, that is, we consider a periodic lattice of unit length in each 
spacetime direction. The discretisation is translation- and reflection 
invariant $\Omega_M^2(m,n)=\Omega_M^2(|| m-n ||)$. The transformation of Euclidian 
actions 
\begin{align}\label{App2}
C\cdot e^{-\beta S'_{M}(\Phi_M)}:=\int  (\prod_m d\tilde{\Phi}_{2M}(m)) e^{-\beta S_{2M}(\tilde{\Phi}_{2M})}e^{-2\kappa \sum_{n\in\mathbb{Z}_M^2}(\Phi_M-\frac{1}{4}\sum_{n'\in\mathbb{Z}^2_{2M},n=\lfloor n'/2\rfloor)}\tilde{\Phi}_{2M})^2}
\end{align}
defines the approximate block spin transformation, where $C$ is some unimportant, $\Phi_M$-independent constant and the exponential on the right hand side 
is called the {\it averaging blocking kernel} that relates the fields $\Phi_M$ on the coarser lattice 
to those $\Phi_{2M}$ on the finer. In the limit $\kappa\rightarrow \infty$ the kernel becomes an exact Dirac-Delta-Distribution, which fixes the new $\Phi_M$ to be an average of all the fields in the old block.

The action is diagonalised using the discrete Fourier transform: 
$\Phi_M(m)=:\frac{1}{M^2}\sum_l e^{ik_M l \cdot m}\hat{\Phi}_M(l)$ and $\Omega^2_M(r)=:\frac{1}{M^2}\sum_l e^{ik_M l\cdot r}\hat{\Omega}_M^2(l)$, with $k_M:=\frac{2\pi}{M}$, $l\in\mathbb{Z}_M^2$. 
We obtain
\begin{align}
\epsilon^2_M\sum_{n,m\in\mathbb{Z}^2_M}\Phi_M(m)\Omega^2_M(m,n)\Phi_M(n)=\epsilon^4_M\sum_{l\in\mathbb{Z}^2_M}\hat{\Phi}_M(l)\hat{\Omega}^2_M(l)\hat{\Phi}_M(-l)
\end{align}
In the literature \cite{BW74} one considers the Hamiltonian $H(\phi)=\frac{1}{2}\int dx \phi (-\Delta) \phi$ and plugs its discretisation straightforwardly into (\ref{App2}).

\subsection{Averaging blocking kernel}
In order to find the fixed point, one studies the generating function $Z(\beta J)$ of $J\in L_M$
\begin{align}
Z(\beta J)&=\frac{1}{Z}\int d\Phi_M e^{-\frac{\beta}{2}\epsilon^2_M\sum_{m,n\in\mathbb{Z}_M^2} \Phi_M(m)\Omega^2_M(m-n)\Phi_M(n)}e^{\beta\epsilon^2_M\sum_{n\in\mathbb{Z}^2_M} J_M(n)\Phi_M(n)}\nonumber\\
&=\frac{1}{Z}\int d\hat{\Phi}_M e^{-\frac{\beta}{2}\epsilon_M^4\sum_{l\in\mathbb{Z}_M^2}\hat{\Phi}_M(l)\hat{\Omega}_M^2(l)\hat{\Phi}_M(-l)+\beta\epsilon^4_M\hat{J}_M(-l)\hat{\Phi}_M(l)}\label{partfunc}
\end{align}
with $Z$ being the partition function and $d\Phi_M$ is the $M^2-$ dimensional 
Lebesgue measure. In a first step, one shifts the variables
\begin{align}
\hat{\Phi}_M(l)=\frac{\hat{J}_M(l)}{\hat{\Omega}^2_M(l)}+\hat{\chi}_M(l)
\end{align}
such that the integral over the $\hat{\chi}_M$ can be computed and cancels the factor $1/Z$:
\begin{align}
Z(\beta J)&=\frac{1}{Z}\int d\hat{\chi}_M \exp\left(-\frac{\beta}{2}\epsilon^4_M\sum_{l\in\mathbb{Z}_M^2}\hat{\chi}_M(l)\hat{\Omega}^2_M(l)\hat{\chi}_M(-l)+\frac{3\beta}{2}\epsilon^4_M\sum_{l\in\mathbb{Z}^2_M}\frac{\hat{J}_M(l)\hat{J}_M(-l)}{\hat{\Omega}^2_M(l)}\right)=\nonumber\\
&=\exp\left(\frac{3}{2}\beta\epsilon_M^4\sum_{l\in\mathbb{Z}^2_M}\frac{\hat{J}_M(l)\hat{J}_M(-l)}{\hat{\Omega}^2_M(l)}\right)
\end{align}
If we expand this expression and (\ref{partfunc}) both to second order in $\hat{J}_M$, we get:
\begin{align}
&\frac{(\beta\epsilon^4_M)^2}{Z}\sum_{l,l'\in\mathbb{Z}^2_M}\hat{J}_M(-l)\hat{J}_M(-l)\int d\hat{\phi}_M 
\hat{\Phi}_M(l)\hat{\Phi}_M(l')\exp\left(-\frac{\beta}{2}\epsilon_M^4\sum_{l_2\in\mathbb{Z}^2_M}\hat{\Phi}_M(l_2)\hat{\Omega}^2_M(l_2)\hat{\Phi}_M(-l_2)\right)\nonumber\\
&=\frac{3}{2}\beta\epsilon_M^4\sum_{l\in\mathbb{Z}^2_M}\hat{J}_M(l)\hat{J}_M(-l)\frac{1}{\hat{\Omega}^2_M(l)}
\end{align}
Since this expression holds for all $\hat{J}_M$ it follows
\begin{align}\label{jcondition}
\frac{1}{Z}\int d\hat{\Phi}_M \hat{\Phi}_M(l)\hat{\Phi}_M(l')e^{-\frac{\beta}{2}\epsilon_M^4\sum_{l_2\in\mathbb{Z}^2_M}\hat{\Phi}_M(l_2)\hat{\Omega}^2_M(l_2)\hat{\Phi}_M(-l_2)}=\left(\frac{3}{2\beta\epsilon_M^4}\right)^{-1}\frac{\delta(l+l')}{\hat{\Omega}^2_M(l)}
\end{align}
This can be used in order to compute the 2-pt-function, which we translate into Fourier space:
\begin{align}
&\langle\Phi_M(n)\Phi_M(n')\rangle=\frac{1}{Z}\int d\Phi_M \Phi_M(n)\Phi_M(n')\exp(-\beta S_M(\Phi_M))\nonumber\\
&=\frac{M^{-4}}{Z}\sum_{l,l'\in\mathbb{Z}^2_M}e^{i(k_Mln+k_Ml'n')}\int d\hat{\Phi}_M\hat{\Phi}_M(l)\hat{\Phi}_M(l')e^{-\frac{\beta}{2}\epsilon_M^4\sum_{l_2\in\mathbb{Z}^2_M}\hat{\Phi}_M(l_2)\hat{\Omega}^2_M(l_2)\hat{\Phi}_M(-l_2)}\overset{(\ref{jcondition})}{=}\nonumber\\
&=\sum_{l,l'\in\mathbb{Z}^2_M}e^{ik_M(ln+l'n')}\delta(l+l')\frac{1}{\hat{\Omega}^2_M(l)}\frac{3}{2\beta}=\nonumber\\
&=\sum_{l\in\mathbb{Z}^2_M}e^{ik_Ml(n-n')}\frac{1}{\hat{\Omega}^2_M(l)}\left(\frac{3}{2\beta}\right)\approx \int_{[0,2\pi]^2}\frac{d^2k}{(2\pi)^2}e^{ik(n-n')}\frac{1}{\epsilon^2_M\hat{\Omega}^2_M(k)}\left(\frac{3}{2\beta}\right)\label{2ptfunc}
\end{align}
The approximation in the last line becomes exact in the {\it continuum limit} $M\to \infty$
in which we may replace $k_M l=2\pi l/M$ by $k\in [0,2\pi]$.

For the renormalisation flow defined by (\ref{App2}), we can compute the 2-pt-function of the coarser lattice in terms of the 2-pt-function on the finer lattice:
\begin{align} \label{complicated}
&\langle \hat{\Phi}_M(n),\hat{\Phi}_M(n')\rangle_M=\\
&=\frac{1}{Z}\int d\Phi_M \left(\Phi_M(n)\Phi_M(n')\right)\exp\left(-\frac{\beta}{2}\epsilon_M^2\sum_{m,m'\in\mathbb{Z}_M}\Phi_{M}(m)(\Omega^2)'_M(m-m')\Phi_M(m')\right)=\nonumber\\
&=\frac{1}{ZC}\int d\Phi_M \left(\Phi_M(n)\Phi_M(n')\right)\int d\Phi_{2M} 
\exp\left(-2\kappa \sum_{m\in\mathbb{Z}^2_M}(\Phi_M(m)-\frac{1}{4}\sum_{m''\in\mathbb{Z}^2_{2M},m=\lfloor m''/2\rfloor}\Phi_{2M}(m''))^2\right)\nonumber\\
&\hspace{25pt}\times \exp\left(-\frac{\beta}{2}\epsilon_{2M}^2 \sum_{m,m'\in\mathbb{Z}^2_{2M}}\Phi_{2M}(m)\Omega^2_{2M}(m-m')\Phi_{2M}(m)\right)=\nonumber\\
&=\frac{1}{ZC}\int d\Phi_{2M} \exp\left(-\frac{\beta}{2}\epsilon^2_{2M}\sum_{m',m''\in\mathbb{Z}_{2M}}\Phi_{2M}(m')\Omega^2_{2M}(m'-m'')\Phi_{2M}(m'')\right)\times\nonumber\\
&\hspace{25pt}\times\int d\Phi_M \left(\Phi_M(n)\Phi_M(n')\right)\exp\left(-2\kappa\sum_{m\in\mathbb{Z}_M} (\Phi_M(m)-\frac{1}{4}\sum_{m'\in\mathbb{Z}_{2M},m=\lfloor m'/2\rfloor}\Phi_{2M}(m'))^2\right)\nonumber
\end{align}
where we used (\ref{App2}) in the second step. Here the appearing integrals over $\Phi_M(\tilde{n})$ are all Gaussians (except for $\tilde{n}=n, n')$
and cancel with the constant $C$. The remaining give back the normalisation $\sqrt{\pi/(2\kappa)}$ twice for $n\not= n'$ and once if $n=n'$. Thus (\ref{complicated}) becomes
\begin{align}
&=\frac{1}{Z}\int d\Phi_{2M} e^{-\beta S_{2M}(\Phi_{2M})}\left(\frac{2\kappa}{\pi}(1-\delta_{n,n'})\sum_{m,m'\in\mathbb{Z}^2_{2M}}\frac{1}{16}\Phi_{2M}(m)\Phi_{2M}(m')\left(\int d\phi e^{-2\kappa \phi^2}\right)^2\right.+\nonumber\\
&\;\;\;\;+\sqrt{\frac{2\kappa}{\pi}}\delta_{n,n'}\frac{1}{16}\sum_{m,m'\in\mathbb{Z}^2_{2M}}\Phi_{2M}(m)\Phi_{2M}(m')\left(\int d\phi e^{-2\kappa\phi^2}\right)
\left.+\sqrt{\frac{2\kappa}{\pi}}\delta_{n,n'} \int d\phi \phi^2 e^{-2\kappa\phi^2}\right)\nonumber\\
&=\frac{1}{16}\sum_{m,m'\in\mathbb{Z}^2_{2M},n=\lfloor m/2 \rfloor,n'=\lfloor m'/2\rfloor}\langle\hat{\Phi}_{2M}(n)\hat{\Phi}_{2M}(n')\rangle_{2M}+\frac{1}{4\kappa}\delta_{n,n'}
\end{align}

Iterating this transformation $j-$times yields
\begin{align}
&\langle\hat{\Phi}_M(n)\hat{\Phi}_M(n')\rangle_M=\\
&=\left(\frac{1}{4}\right)^{2j} \sum_{m,m'\in\mathbb{Z}^2_{2^jM},n=\lfloor m/2^j\rfloor,n'=\lfloor m'/2^j\rfloor}\langle\hat{\Phi}_{2^j M}(n)\hat{\Phi}_{2^j M}(n')\rangle_{2^jM}+\frac{1}{4\kappa}(1+\frac{1}{4}+\frac{1}{4^2}+...+\frac{1}{4^{j-1}})\delta_{n,n'}\nonumber
\end{align}
Simultaneous with the limit $j\rightarrow\infty$, we consider the original lattice to become infinitely fine, such that the summations in the first term on the right-hand side go over to integrals for which we must absorb the factor $2^{-j}$.\\
Moreover, it is assumed safe in \cite{HN93} to perform the limit of the 2-pt function separately and plug in the standard $\frac{1}{p^2}$ propagator for the infinitely fine lattice. Following their strategy, we arrive for large $j$ at
\begin{align}\label{continuum2ptft}
\langle \hat{\Phi}_M(n)\hat{\Phi}_M(n')\rangle\approx\int_{-1/2}^{1/2}d^2x\int_{-1/2}^{1/2}d^2x'\left(\int_{-\infty}^{\infty}\frac{d^2p}{(2\pi)}\frac{e^{ip\cdot(n+x-n'-x')2\epsilon_M}}{p^2}\right)+\frac{1}{3\kappa}\delta_{n,n'}
\end{align}
Now we compare (\ref{continuum2ptft}) with (\ref{2ptfunc}) - which was the 2-pt function at a fixed-point - by diving the $p_i$ integration into a summation of the integer $l_i$ and an integration over $k_i$, i.e. $p=k+2\pi l $ such that:
\begin{align}
&\int_{-\pi}^{\pi}\frac{d^2k}{(2\pi)^2} e^{ik(n-n')}\frac{1}{\hat{\Omega}^2_M(k)}\frac{3}{2\beta\epsilon^2_M}=\nonumber\\
&=\int_{0}^{2\pi}\frac{d^2k}{(2\pi)^2}\sum_{l\in\mathbb{Z}^2}\frac{1}{(k+2\pi l)^2}\int_{1/2}^{-1/2}dx\int_{1/2}^{-1/2}dx' e^{i(k+2\pi l)(n+x-n'-x')}+\int_{-\pi}^{\pi}\frac{d^2k}{2\pi}e^{ik(n-n')}\frac{1}{3\kappa}
\end{align}
It follows:
\begin{align}
\frac{1}{\epsilon^2_M\hat{\Omega}^2_M(k)}&=\frac{2\beta}{9\kappa}+\sum_{l\in\mathbb{Z}^2}\frac{2\beta/3}{(k+2\pi l)^2}e^{i2\pi l(n-n')}\frac{(-1)}{(k/2+\pi l)^2}\frac{[e^{i(k+2\pi l)x}]^{1/2}_{-1/2}}{2i}\frac{[e^{-i(k+2\pi l)x'}]^{1/2}_{-1/2}}{2i}=\nonumber\\
&=\sum_{l\in\mathbb{Z}^2}\frac{2\beta/3}{(k+2\pi l)^2}\prod^1_{\mu=0}\frac{\sin(k_{\mu}/2+\pi l_{\mu})^2}{(k_{\mu}/2+\pi l_{\mu})^2}+\frac{2\beta}{3\kappa}\label{approxaveragefixpoint}
\end{align}
This is the final expression for the covariance at the fixed point as found in the literature.\\

\subsection{Deleting Blocking kernel}
In this subsection we repeat the analysis of the previous subsection 
for the deleting kernel
\begin{align}\label{deletingkernel}
e^{-2\kappa \sum_{m'\in\mathbb{Z}^2_{M}}(\Phi_M(m)-\Phi_{2M}(2m))^2}
\end{align}
In the main text it was applied only in the spatial direction, but in order to relate with the literature we use it here in the spacetime sense.

We compute the flow defined by this kernel by looking again at the 2-pt-function on a coarse lattice in terms of the 2-pt-function on the finer lattice:
\begin{align} \label{complicatedalso}
&\langle \hat{\Phi}_M(n)\hat{\Phi}_M(n')\rangle_M=\nonumber\\
&=\frac{1}{Z}\int d\Phi_M \left(\Phi_M(n)\Phi_M(n')\right)\exp\left(-\frac{\beta}{2}\epsilon_M^2\sum_{m,m'\in\mathbb{Z}^2_M} \Phi_M(m)
(\Omega^2)'_M(m-m')\Phi_M(m')\right)=\nonumber\\
&=\frac{1}{Z\cdot C}\int d\Phi_M \left(\Phi_M(n)\Phi_M(n')\right)\int d\Phi_{2M}\times\nonumber\\
&\hspace{8pt}\times \exp\left(-2\kappa \sum_{m\in\mathbb{Z}^2_M} (\Phi_M(m)-\Phi_{2M}(2m))^2-\frac{\beta}{2\epsilon_M}\sum_{m,m'\in\mathbb{Z}^2_{2M}}\Phi_{2M}(m)\Omega^2_{2M}(m-m')\Phi_{2M}(m')\right)\nonumber\\
&=\frac{1}{Z\cdot C}\int d\Phi_{2M} \exp\left(-\frac{\beta}{2\epsilon_M}\sum_{m,m'\in\mathbb{Z}^2_{2M}}\Phi_{2M}(m)\Omega^2_{2M}(m-m')\Phi_{2M}(m')\right)\times\nonumber\\
&\hspace{10pt}\times\int d\Phi_{M} \left(\Phi_{M}(n)\Phi_{M}(n')\right) \exp\left(-2\kappa \sum_{m\in\mathbb{Z}^2_M}(\Phi_M(m)-\Phi_{2M}(2m))^2\right)
\end{align}
where we used (\ref{deletingkernel}) in the second step. The integrals over $\Phi_{2M}(m)$ are all Gaussian, except for $m=n,n'$, and cancel with the constant $C$. The remaining integrals return two or one factors of the normalisation $\sqrt{\pi/(2\kappa)}$ and (\ref{complicatedalso}) becomes
\begin{align}
&=\frac{1}{\left(\frac{\pi}{2\kappa}\right)Z}\int d\Phi_{2M} e^{-\beta S_{2M}(\Phi_{2M})}\left(
(1-\delta_{n,n'})\prod_{i=n,n'}\int d\Phi_{M}(i) \right.\times\nonumber\\
&\hspace{20pt}\times\left.\left(\Phi_{M}(i)-\Phi_{2M}(2i)+\Phi_{2M}(2i)\right)e^{-2\kappa\sum_{i=n,n'}(\Phi_M(i)-\Phi_{2M}(2i))^2}+\right.\nonumber\\
&\hspace{10pt}+\left.\delta_{n,n'}\sqrt{\frac{\pi}{2\kappa}}\int d\Phi_M(n)
\left(\Phi_{M}(n)-\Phi_{2M}(2n)+\Phi_{2M}(2n)\right)^2e^{-2\kappa (\Phi_M(n)-\Phi_{2M}(2n))^2}
\right)\nonumber\\
&=\langle \hat{\Phi}(2n)\hat{\Phi}(2n')\rangle_{2M}+\frac{1}{4\kappa}\delta_{n,n'}
\end{align}
After $j$ steps of iteration
\begin{align}
\langle\hat{\Phi}_M(n)\hat{\Phi}_M(n')\rangle^{(j)} =\langle \hat{\Phi}_{2^j M}(2^jn)
\hat{\Phi}_{2^j M}(2^jn') \rangle+\frac{j}{4\kappa}\delta_{n,n'}
\end{align}
In the limit $j\rightarrow \infty$ the last term is problematic unless we 
take first the limit $\kappa\rightarrow \infty$.
Note that $\Phi_{2^j M}(2^j n)=\Phi(2^j n \epsilon_{2^j M})=\Phi(n \epsilon_M)$ in terms of 
the continuum field. 
For the 2-pt-function in the continuum we take the standard $1/p^2$ propagator. In total
\begin{align}
\langle\hat{\Phi}_M(n)\hat{\Phi}_M(n')\rangle_M=\int_{-\infty}^{\infty}\frac{dp}{(2\pi)}\frac{1}{p^2}e^{ip(n+\frac{1}{2}-n'-\frac{1}{2})(2\epsilon_M)}
\end{align}
Now we compare this to (\ref{2ptfunc}) by diving the $p$ integration again into a summation of the integer $l$ and an integration over $k$, i.e. $p= k+2\pi l$, whence 
\begin{align}
\int^{\pi}_{\pi}\frac{dk}{2\pi}e^{ik(n-n')}\frac{1}{\hat{\Omega}^2_{M}(k)}\left(\frac{3}{2\beta\epsilon^2_M}\right)&=\int^{\pi}_{-\pi}\frac{dk}{2\pi}\sum_{l\in\mathbb{Z}^2_M}\frac{2\epsilon_M}{(k+2\pi l)^2}e^{i(k+2\pi l) (n-n')}\\
\Rightarrow \frac{1}{\epsilon^2_M\hat{\Omega}^2_M}&=\sum_{l\in\mathbb{Z}^2_M}\frac{2\beta/3}{(k+2\pi l)^2}
\end{align}\\

\end{document}